\documentclass[acmtog,nonacm]{acmart}
\acmSubmissionID{12345}
\usepackage{multirow}
\usepackage{booktabs} 
\usepackage{natbib}
\usepackage{graphicx}
\usepackage{float}
\usepackage{enumitem}
\usepackage{bbding}
\usepackage{soul}
\usepackage{bm}
\usepackage{xspace}
\usepackage{array}
\usepackage{multirow}
\usepackage{hyperref}
\usepackage[ruled]{algorithm2e} 

\SetAlFnt{\small}
\SetAlCapFnt{\small}
\SetAlCapNameFnt{\small}
\SetAlCapHSkip{0pt}

\newcommand{\red}[1]{{\color{red}#1}}

\begin{document}
\title{DreamFace: Progressive Generation of Animatable 3D Faces under Text Guidance}

\renewcommand\shortauthors{XXX et al}
\newcommand{\eqtr}{\mathrel{\raisebox{-0.1ex}{%
\scalebox{0.8}[0.6]{$\vartriangle$}}}}

\author{Longwen Zhang}\authornote{Equal contributions.}
\orcid{0000-0001-8508-3359}
\affiliation{%
 \institution{ShanghaiTech University and Deemos Technology Co., Ltd.}
 \city{Shanghai}
 \country{China}}
\email{zhanglw2@shanghaitech.edu.cn}

\author{Qiwei Qiu}\authornotemark[1]
\affiliation{%
 \institution{ShanghaiTech University and Deemos Technology Co., Ltd.}
 \city{Shanghai}
 \country{China}}
\email{qiuqw@shanghaitech.edu.cn}

\author{Hongyang Lin}\authornotemark[1]
\affiliation{%
 \institution{ShanghaiTech University and Deemos Technology Co., Ltd.}
 \city{Shanghai}
 \country{China}}
\email{linhy@shanghaitech.edu.cn}

\author{Qixuan Zhang}
\affiliation{%
 \institution{ShanghaiTech University and Deemos Technology Co., Ltd.}
 \city{Shanghai}
 \country{China}}
\email{zhangqx1@shanghaitech.edu.cn}

\author{Cheng Shi}
\affiliation{%
 \institution{ShanghaiTech University}
 \city{Shanghai}
 \country{China}}
\email{shicheng2022@shanghaitech.edu.cn}

\author{Wei Yang}
\affiliation{%
 \institution{Huazhong University of Science and Technology}
 \city{Wuhan}
 \country{China}}
\email{weiyangcs@hust.edu.cn}

\author{Ye Shi}
\affiliation{%
 \institution{ShanghaiTech University}
 \city{Shanghai}
 \country{China}}
\email{shiye@shanghaitech.edu.cn}

\author{Sibei Yang}
\affiliation{%
 \institution{ShanghaiTech University}
 \city{Shanghai}
 \country{China}}
\email{yangsb@shanghaitech.edu.cn}

\author{Lan Xu}
\affiliation{%
 \institution{ShanghaiTech University and Shanghai Engineering Research Center of Intelligent Vision and Imaging}
 \city{Shanghai}
 \country{China}}
\email{xulan1@shanghaitech.edu.cn}

\author{Jingyi Yu}
\affiliation{%
 \institution{ShanghaiTech University and Shanghai Engineering Research Center of Intelligent Vision and Imaging}
 \city{Shanghai}
 \country{China}}
\email{yujingyi@shanghaitech.edu.cn}

\renewcommand\shortauthors{Zhang, L. et al}

\begin{abstract}
Emerging Metaverse applications demand accessible, accurate and easy-to-use tools for 3D digital human creations in order to depict different cultures and societies as if in the physical world. Recent large-scale vision-language advances pave the way for novices to conveniently customize 3D content. However, the generated CG-friendly assets still cannot represent the desired facial traits for human characteristics. 
In this paper, we present DreamFace, a progressive scheme to generate personalized 3D faces under text guidance. It enables layman users to naturally customize 3D facial assets that are compatible with CG pipelines, with desired shapes, textures and fine-grained animation capabilities. 
From a text input to describe the facial traits, we first introduce a coarse-to-fine scheme to generate the neutral facial geometry with a unified topology. We employ a selection strategy in the CLIP embedding space to generate coarse geometry, and subsequently optimize both the detailed displacements and normals using Score Distillation Sampling (SDS) from the generic Latent Diffusion Model (LDM).
Then, for neutral appearance generation, we introduce a dual-path mechanism, which combines the generic LDM with a novel texture LDM to ensure both the diversity and textural specification in the UV space. We also employ a two-stage optimization to perform SDS in both the latent and image spaces to significantly provide compact priors for fine-grained synthesis. It also enables learning the mapping from the compact latent space into physically-based textures (diffuse albedo, specular intensity, normal maps, etc.).
Our generated neutral assets naturally support blendshapes-based facial animations, thanks to the unified geometric topology. We further improve the animation ability with personalized deformation characteristics.
To this end, we learn the universal expression prior in a latent space with neutral asset conditioning using the cross-identity hypernetwork, we subsequently train a neural facial tracker from video input space into the pre-trained expression space for personalized fine-grained animation.
%
Extensive qualitative and quantitative experiments validate the effectiveness and generalizability of DreamFace.
Notably, DreamFace can generate realistic 3D facial assets with physically-based rendering quality and rich animation ability from video footage, even for fashion icons or exotic characters in cartoons and fiction movies.

\end{abstract}

%



%
%


\begin{teaserfigure}
    \setlength{\abovecaptionskip}{3pt}
    \centering
    \includegraphics[width=1\textwidth]{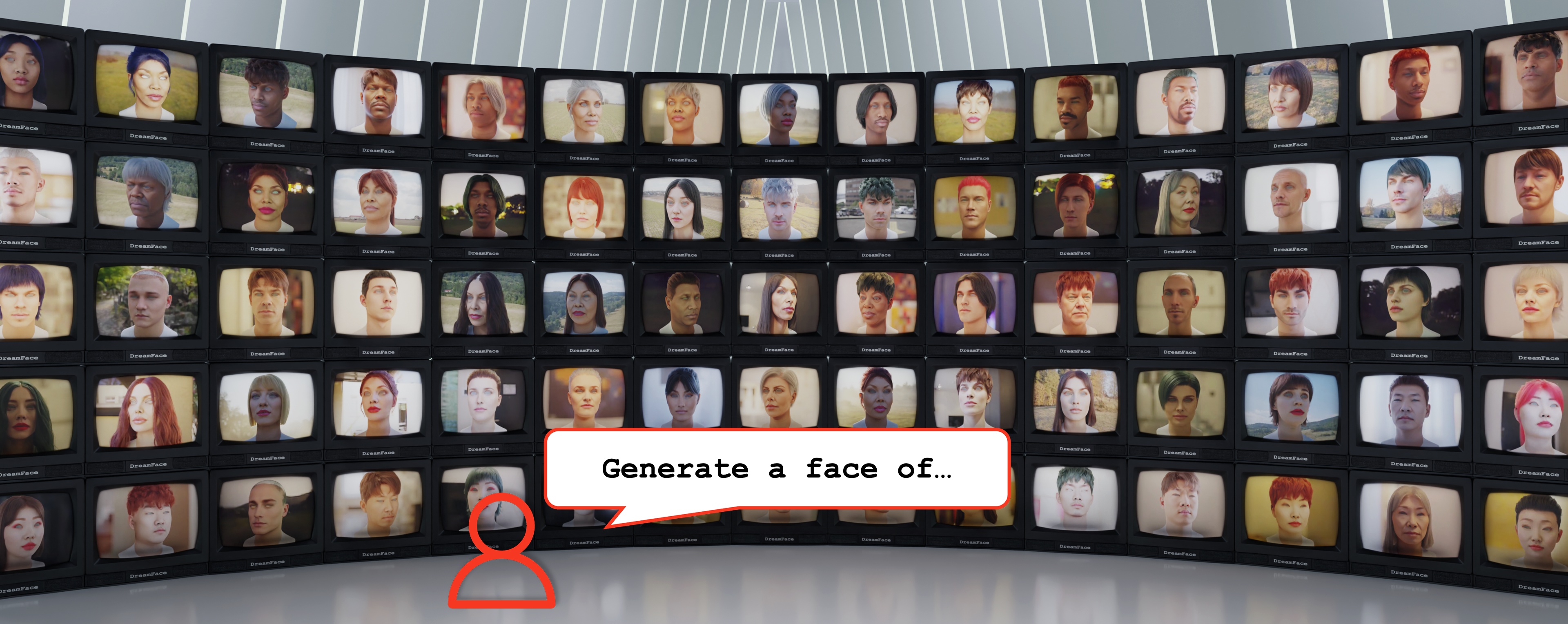}
    \caption{
{
DreamFace generates personalized 3D physically-based facial assets under text guidance, which are compatible with the existing CG pipeline, with desired shapes, textures, and fine-grained animations for realistic rendering. Go to DreamFace project page  \red{\url{https://sites.google.com/view/dreamface}}, watch our video at \red{\url{https://youtu.be/yCuvzgGMvPM}} and experience DreamFace online at \red{\url{https://hyperhuman.top}} !
}
    }
    \label{fig:teaser}
\end{teaserfigure}

\maketitle

\section{Introduction}


Digital human face models should equally reflect the diversity of the human experience in order to depict different cultures, societies, and environments that make up our physical world. A successful 3D digital human creation tool will hence allow users to create and customize models that follow their own identities and physical characteristics. These include skin color, hair types, facial shapes, expressions and appearance with rich facial traits. In addition to being realistic and believable, the produced assets also need to match specific themes or concepts, e.g., from as complicated as the movie script and literature, or as simple as textual descriptions by novice users. It has been in high demand for a variety of applications for feature films, game productions, and most recently, immersive experiences in the Metaverse. 

However, creating these believable 3D human characters is not for anyone. Early attempts~\cite{2000lightstage, EmilyProject} generally require expensive apparatus and immense artistic expertise, and hence are limited to celebrities for feature film productions. It has been a long journey for the graphics community to democratize the accessible use of 3D facial assets to the mass crowd, equipped with powerful neural generative techniques, from variational autoencoders (VAEs)~\cite{kingma2014vae}, generative adversarial networks (GANs)~\cite{goodfellow2020ganorigin} to the latest Diffusion Models~\cite{ho2020ddpm}. 
Seminal generation approaches such as StyleGAN2~\cite{tero2020stylegan2} have successfully produced highly realistic 2D facial rendering almost indistinguishable from real photos.
Various methods~\cite{orel2022styleSDF, Chan2021eg3d, deng2022gram} further combine 2D GANs with 3D implicit representations, achieving 3D-aware facial synthesis with detailed hairstyles, textures or expressions. Yet, these rendering-based schemes are still difficult to integrate seamlessly into the existing CG production pipeline. 
A few recent works~\cite{li2020Dynamic,li2020learningformation} combine production-ready facial assets into the VAE or GAN-based generation framework. But they still are not able to produce a high degree of diversity, partially due to the limited training set of 3D or paired data and the limited flexibility of the generation process. Hence, the vision of creating facial assets from novices' text prompt remains far-reaching.
Fortunately, recent large-scale vision-language models~\cite{rombach2022stablediffusion, radford2021clip} pave the way to lower the barrier to entry for novices further.  
Using the advanced diffusion models, recent attempts have achieved huge success for zero-shot 2D image content creation from text prompts~\cite{ramesh2022dalle2,saharia2022imagen}, or even for 3D generation\cite{jain2022dreamfields,poole2022dreamfusion,lin2022magic3d,metzer2022latent-nerf}.
A few works utilize the pre-trained CLIP model~\cite{radford2021clip} to generate animatable full-body avatars~\cite{hone2022avatarclip} or facial texture maps~\cite{aneja2022clipface}.
Despite the diversity, the generated assets from the above methods still cannot represent the desired facial traits for human characteristics, e.g., lacking detailed facial geometry, fine-grained animation or physically-based textures.

In this paper, we present DreamFace, a progressive scheme to generate personalized 3D faces with text guidance. As shown in Fig.~\ref{fig:teaser}, with only prompt controls, DreamFace enables layman users to customize 3D facial assets with the desired shape and physically-based textures, as well as empowered animation capabilities. The resulting assets (mesh, texture, normal maps, etc.) are also compatible with the existing graphics production pipeline, significantly democratizing the use of generated facial assets. In particular, we demonstrate that DreamFace not only benefits the CG production industry but also incentivizes numerous innovative applications.

At the core of our approach is the organic integration of the recent large-scale vision-language advances with dynamic physically-based facial assets. 
For high-quality and more controllable generation, we design DreamFace in a progressive framework, which consists of three sequential modules: geometry generation, physically-based texture diffusion, and animation empowerment.
From a text input to describe the facial traits of desired avatar, we first introduce a coarse-to-fine scheme to generate the corresponding neutral facial geometry with the topology structure from ICT-FaceKit~\cite{li2020learningformation}. 
In the coarse stage, we select the optimal coarse geometry from a diverse candidate pool randomly pre-sampled from the shape space of ICT-FaceKit, by comparing relative matching scores between the prompt and the rendered geometry images in the CLIP embedding space. 
We then perform fine-grained detail carving on top of the coarse geometry regarding vertex displacements and detailed normal maps in the tangent space. 
Analogous to DreamFusion~\cite{poole2022dreamfusion}, we learn the detailed displacements and normal maps using the Score Distillation Sampling technique (SDS) to compute gradients from the generic Latent Diffusion Model (LDM), i.e., Stable Diffsuion~\cite{rombach2022stablediffusion}.
Then, our physically-based appearance generation aims to predict the neural facial assets that are consistent with both the predicted geometry and text prompt. 
Specifically, we adopt a dual-path mechanism to utilize two kinds of diffusion models: one generic LDM for diverse generation ability from general prompt inputs and a novel texture LDM to ensure the textural specifications in the UV space. For training our texture LDM, we augment existing UV texture datasets with our physically-based one and subsequently adopt a prompt tuning strategy to compensate for the difference between various datasets. For efficient appearance generation, we further introduce a two-stage optimization to perform Score Distillation Sampling (SDS) in both the latent and image spaces, where the latent one significantly provides compact priors for fine-grained synthesis. Note that both SDS processes are enhanced with the generic and texture LDMs through our dual-path design.
To obtain physically-based assets, we further learn the mapping from the compact latent space into diffuse albedo, specular intensity, and normal maps, followed by a super-resolution module to generate 4K textures for high-quality rendering.

For animating the generated neutral geometry and appearance, the brute-force approach would adopt the default blendshapes provided by the parametric model ICT-FaceKit~\cite{li2020learningformation}, since our assets share the same geometric topology with ICT-FaceKit.
One could utilize existing face trackers to obtain the corresponding expression parameters from video footage so as to seamlessly animate our facial assets. 
We further introduce an enhanced animation scheme accompanied by our neutral assets to maintain more personalized motion characteristics than the general blendshapes. 
We first follow the recent work~\cite{phonescan} to adopt the cross-identity hypernetwork, so as to learn a universal prior for modeling the expression space of the generated facial assets. We adopt the U-Net architecture to transform the neutral asset (in terms of geometry images) into the generated facial mesh under various expressions.
With this universal prior with neutral asset conditioning, we train a video facial tracker as an encoder from the image space into the pre-trained expression space, enabling video-based personalized animation.
Finally, we showcase the capability of DreamFace for generating realistic 3D facial assets with rich animation and physically-based rendering quality, even for fashion icons or exotic characters in cartoons and fiction movies.
With our DreamFace, only through textural guidance, even novices can naturally create the human characters they have in mind and easily customize the creations for novel effects like aging and virtual makeup, and even further animate the creations using in-the-wild video footage.


To summarize, our main contributions include:
\begin{itemize} 
    \setlength\itemsep{0em} 
	
    \item {We present DreamFace, a novel generation scheme to bridge recent vision-language models with animatable and physically-based facial assets, with progressive learning to disentangle geometry, appearance and animation ability.}
	
    \item {We introduce a dual-path appearance generation design to combine a novel texture diffusion model with the pre-trained one, equipped with a two-stage optimization in both the latent and image spaces.}
	
    \item {We demonstrate the animation ability of the generated facial assets using blendshapes or an empowered personalized scheme, and further showcase the applications of DreamFace to design human characters naturally.}
 
\end{itemize} 

\section{Related Works}





\paragraph{Face modeling and generation}
The generation of 3D faces has seen rapid progress in recent years. Many research efforts have attempted to generate realistic facial modeling, which draws a solid foundation for generation research. 
Parametric models using Principal Component Analysis (PCA) were first used to express facial components containing geometry and textures, especially 3DMM model~\cite{blanz19993dmmorigin}. More work extended the expressiveness of  parametric models to represent the facial details faithfully \cite{cao2013facewarehouse, li2017flame, ranjan2018coma}. 
GAN~\cite{goodfellow2020ganorigin} methods demonstrated their strong generative ability to produce high-fidelity results. Seminal generation approaches such as StyleGAN2~\cite{tero2020stylegan2} generate the face image of a vast abundance. 
Extend to 3D generation, a series of methods~\cite{liu20223dfmgan, tewari2020stylerig, orel2022styleSDF, Chan2021eg3d, deng2022gram} focused on embedding 3D priors within the GAN framework to generate 3D faces consistent with known attributes. 
Such generative methods model 3D faces with varieties of representation, including explicit representation based on mesh ~\cite{liao2020towards, gecer2021fastganfit, abrevaya2019decoupled, lattas2021avatarme++, li2020learningformation} and implicit representation based on SDF, such as StyleSDF~\cite{orel2022styleSDF}, or based on NeRF~\cite{mildenhall2021nerf}, like HeadNeRF~\cite{hone2022headnerf} and MofaNeRF~\cite{zhuang2022mofanerf}. 
Although implicit representation methods achieve remarkable detail quality and visual results, they are hard to be integrated into the existing CG production pipeline. 
On the other hand, even though explicit representation methods enjoy good compatibility, the over-smoothed geometry cannot faithfully reproduce detailed features. 
%
In this paper, we design DreamFace, a highly compatible scheme that can generate highly realistic and detailed facial assets. Specifically, DreamFace applies the explicit face mesh with the same geometric topology as ICT-FaceKit~\cite{li2020learningformation} to utilize displacement and tangent space normal maps to achieve fine-grained rendering.

\begin{figure*}[t]
    \centering
    \includegraphics[width=\linewidth]{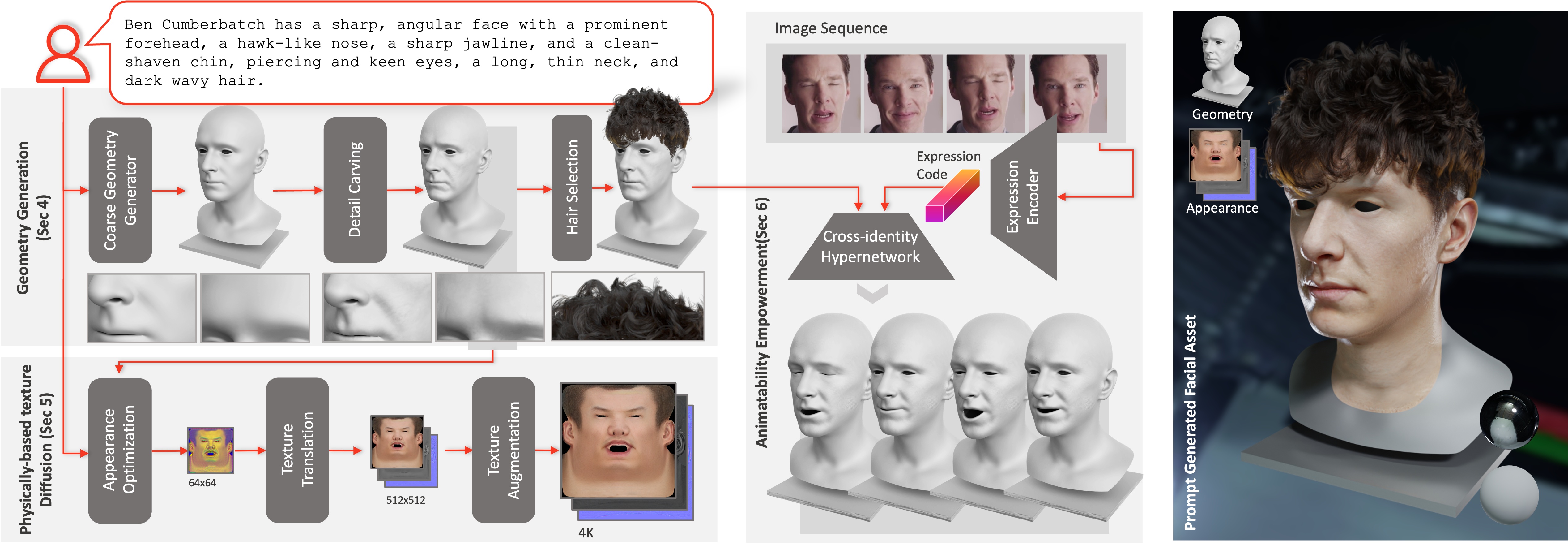}
    \caption{
    The overview of DreamFace. Our pipeline mainly includes three modules, including geometry generation (Sec.~\ref{sec:geometry}), physically-based texture diffusion (Sec.~\ref{sec:appearance}), and animatability empowerment (Sec.~\ref{sec:animation}).
Given textual guidance, DreamFace is able to generate facial assets that closely resemble the described characteristics in terms of shape and appearance. Our approach is consistent with industry standards in computer graphics production and is able to achieve photo-realistic results when driven and rendered.
    }
    \label{fig:overview}
\end{figure*}

\paragraph{Texture generation}
Generating faces with mesh representation always requires high-quality facial textures. There is a large corpus of research works in the field of generative models for UV textures based on self-supervised methods~\cite{gecer2019ganfit, gecer2021fastganfit, lee2020styleuv, fukamizu20193dgan1, gecer2020tbgan}. 
Though these self-supervised methods reached relatively satisfactory results, their performance is still far from ultra-realistic rendering. In contrast, supervised methods~\cite{Lattas2020avatarme, lattas2021avatarme++, li2020learningformation, bao2021hifi3dface} adopted image-to-image translation frameworks \cite{isola2017pix2pix, wang2018pix2pixhd} to generate another reflectance textures from a single albedo map, ensuring high-level correspondence between the generated textures. However, they still have difficulty assembling a large-scale dataset that always requires a photometric capture system~\cite{ma2007rapidacquistion, ghosh2011polarizedgradient}. Different texture standards and data quality further hinder joint training across different datasets. In addition, most of these methods are based on GANs that are neither scalable nor stable on text-conditional generation tasks. 
Recent advances show the unprecedented generative abilities and compositional power of Diffusion models~\cite{sohl2015diffusion, ho2020ddpm}, which have not yet been developed in texture generation. 
To this end, we propose to train the diffusion model for producing high-quality physically-based textures with a novel prompt tuning strategy to bridge the uneven quality across all training data.

\paragraph{Text to 3D generation}
In recent years, there has been a great deal of interest in text-driven image generation
~\cite{mansimov16_text2image, reed2016gantextimage}. Leverage on the powerful vision-language representations enriched by CLIP~\cite{radford2021clip}, DALL-E2~\cite{ramesh2022dalle2} further demonstrates zero-shot text-driven image synthesis capabilities by excessively scaling up training data size. 
 Many works, such as VQGAN-CLIP~\cite{crowson2022vqganclip} and GLIDE~\cite{nichol2021glide}, guided the generative image models by embedding the distance with the latent code of text prompts encoded by the CLIP model. StyleCLIP~\cite{patashnik2021styleclip} followed by StyleGAN-Nada~\cite{gal2022stylegan-nada} combined the generative power of StyleGAN and the expressive power of CLIP, enabling the image generation to have remarkable diversity from various domains. The state-of-the-art text-to-image generation method~\cite{rombach2022stablediffusion} uses the Latent Diffusion Model that shifts the diffusion process to the latent space using pretrained autoencoders. This approach makes the training more stable and feasible to train on larger datasets to achieve a more remarkable richness. 
In contrast to the well-developed 2D image generation task, 3D generation needs more effort to generate high-quality and divisible 3D models. 
\citet{sanghi2022clipforge} attempted to directly generate 3D objects by training a autoencoder with 3D representations.
Further combination of the vision-language model with the rendered image from differentiable rendering on mesh representation~\cite{michel2022text2mesh, khalid2022clipmesh, chen2022tango} or NeRF representation~\cite{hone2022avatarclip, jain2022dreamfields} gives more progress to the generation and editing of 3D objects.
DreamFusion~\cite{poole2022dreamfusion} pioneered the introduction of the diffusion model for the supervision of generated objects but suffered for a long generation time.
Magic3D~\cite{lin2022magic3d} accelerates the DreamFusion~\cite{poole2022dreamfusion} by optimizing the initial coarse model generated by low-resolution diffusion prior, using a sparse 3D hash grid structure~\cite{thomas2022instantngp}.
Latent-NeRF~\cite{metzer2022latent-nerf} utilized LDM to directly optimize NeRF on latent space and proved the effectiveness of optimizing latent textures on the mesh. 
The above methods either use CLIP or the Diffusion model to generate stunning results but still suffer from Two-faces Janus problems, resulting in insufficient quality.
To handle this problem, we develop a novel dual-path optimization scheme that leverages the learned texture prior to guiding the Score Distillation Sampling and still retains powerful text-guided generation capability.

\paragraph{Facial Animation}
Unlike implicit representation, explicit representation supports generating expressions directly by generic expression blendshapes. 
Some face-tracking techniques~\cite{fyffe2015scanopti, feng2021deca, blendshapeanimation,arkit} can provide a lightweight capture solution for facial animation with a single RGB camera input.
To further enhance the inaccurate result of generic facial animation, many research efforts made huge progress. \citet{laine2017di4d} trained a characteristic neural network with registered 4D scans. 
\citet{lombardi2018dam} encodes geometry and view-dependent texture information with a VAE framework that enables video-driven animation from VR Headset cameras. 
To allow for cross-identity retargeting, \citet{moser2021semi} applied image-to-image translation and extracted a common representation between the input video and rendered CG sequence to predict blendshape weights.
The state-of-the-art neural animation method~\cite{zhang2022npfa} proposed a production-level animation pipeline that generates high-quality facial geometry for person-specific performance.
A very recent work \citet{phonescan} proposed a Universal Prior Model (UPM), which can produce personalized expressions from unseen identities. 
Inspired by this work, we train a cross-identity hypernetwork together with an image expression encoder to enable personalized animation for our generated facial assets by novice users.

\section{Overview}
Here we introduce DreamFace, a progressive generation scheme to marry large-scale vision-language models with personalized facial assets that are compatible with CG engines.
As shown in Fig.~\ref{fig:overview}, from only prompt controls, our DreamFace disentangles the generation framework into three cooperated modules, named geometry generation, physically-based texture diffusion, and animation empowerment, respectively.

\begin{figure*}[t]
    \centering
    \includegraphics[width=\linewidth]{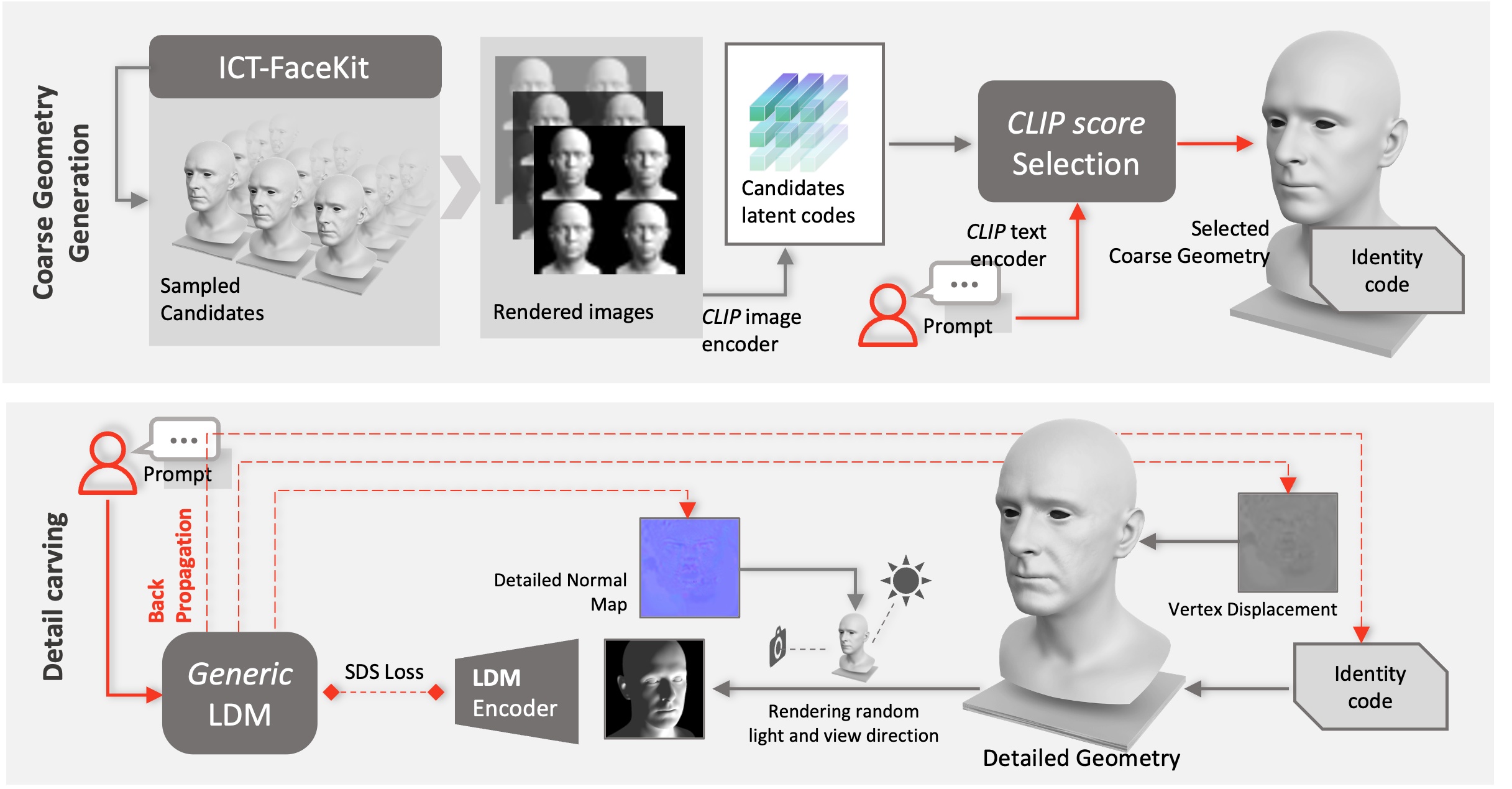}
    \caption{
Geometry generation pipeline. Given the input prompt, we utilize the CLIP model to select the coarse geometry candidates with the highest matching score. Next, we employ a generic LDM to perform SDS on the rendered images under random view and lighting conditions. This allows us to add facial details to the coarse geometry via vertex displacement and detailed normal map, resulting in a highly detailed geometry.
    }
    \label{fig:geometry}
\end{figure*}

For geometry generation, we adopt a coarse-to-fine scheme to obtain neutral geometry with fine-grained facial traits and topology structure from the parametric model ICT-FaceKit~\cite{li2020learningformation} (Sec.~\ref{sec:geometry}). In the coarse stage, we select an optimal one from a diverse candidate pool pre-sampled from ICT-FaceKit by calculating the matching scores based on CLIP model~\cite{radford2021clip}. Then, we perform fine-grained detail carving on top of the coarse geometry by applying the Score Distillation Sampling technique~\cite{poole2022dreamfusion} to optimize both the detailed displacements and normal maps in the tangent space. We also adopt an efficient prompt-based hair selection to increase realism. 

We further generate a physically-based appearance consistent with both the predicted neutral geometry and text prompt (Sec.~\ref{sec:appearance}). We introduce a dual-path mechanism with two latent diffusion models (LDMs), including a generic LDM for diversifying the generated results from arbitrary prompt inputs, as well as a novel texture LDM in the UV space. The texture LDM is trained using augmented UV texture datasets with physically-based ones, as well as prompt tuning for the compensation of data variance. We also introduce a two-stage optimization to perform the Score Distillation Sampling (SDS) in both the latent and image spaces, which are enhanced through our dual-path design.  It enables well-sculpted latent space with compact priors for efficient and fine-grained synthesis. Finally, we learn the mapping from the latent space into physically-based assets and subsequently adopt a super-resolution to generate 4K textures.

For animation, our generated neutral assets naturally support existing facial trackers to obtain the expression parameters, benefiting from the by-default blendshapes and topology from ICT-FaceKit. We further empower the animation ability of our assets to model personalized deformation characteristics (Sec.~\ref{sec:animation}).
We employ the cross-identity hypernetwork~\cite{phonescan} to learn the universal expression prior in a latent space, where a U-Net architecture is adopted to transform the neutral asset (as geometry images) into the generated facial meshes with diverse expressions. On top of such expression prior with neutral asset conditioning, we trained a neural video facial tracker, as an encoder from the image space into the pre-trained expression space, achieving video-based personalized fine-grained animation.

\section{Geometry Generation}
\label{sec:geometry}

From a user prompt $\mathcal{P}$ that describes the facial characteristics, we first introduce a coarse-to-fine scheme to generate the corresponding neutral facial geometry with the topology structure from ICT-FaceKit, which consists of 14062 vertices and 28068 faces. 
We propose a prompt-based selection framework that chooses the optimal coarse geometry from diverse candidates randomly sampled from ICT-FaceKit shape space, which has the best CLIP matching score with user prompt. 
We then perform fine-grained detail carving on top of the coarse geometry in terms of vertex displacements and detailed normal maps in the tangent space. Similar to DreamFusion \cite{poole2022dreamfusion}, We rely on the Score Distillation Sampling (SDS) loss of the pretrained generic LDM, i.e., Stable Diffusion \cite{rombach2022stablediffusion}, for guiding the details carving. 
We also use the pre-trained LDM image autoencoder from Stable Diffusion with $\mathcal{E}$ as encoder and $\mathcal{D}$ as decoder, which converts a 512$\times$512 image to 64$\times$64 latent code and back.
In the following, we first describe the process of generating coarse geometry and then introduce the facial details carving process based on the coarse geometry.

\subsection{Coarse geometry generation}


As depicted in Fig.~\ref{fig:geometry}, the coarse geometry generation process involves randomly sampling candidates from the shape space of ICT-FaceKit~\cite{li2020learningformation}. 
We then render the front and left/right 3/4 views of the selected face geometry. And then, we use the CLIP model~\cite{radford2021clip} to extract features from the images and select the one that best matches the input user prompt. We use the geometry corresponding to the best match as our coarse head mesh.
%

For sampling candidates from ICT-FaceKit, we obtain a shape space with $|\bm{\beta}|=100$ basis. The formulate of a certain shape from the shape space is:
\begin{equation}
    \mathbf{T}=T(\bm{\beta})=\mathbf{\bar T}+\sum_i\beta_i\mathbf{S}_i,
\end{equation}
where $\mathbf{\bar T}$ is the mean face, $\mathbf{S}_i$ is the shape components, and $\mathbf{T}$ are corresponding generated head mesh. 
The coarse geometry candidates are sampled from the shape space via choose ${\bm \beta}$ from a multivariate normal distribution $\bm{\beta} \sim \mathcal{N}(\mathbf{0}, \mathbf{1})$,
Then, the sampled candidates share a similar distribution that covers the span of the possible facial shapes in ICT-FaceKit. 
%

To find the best match from candidate geometries, we first render the front and left/right 3/4 views of the selected face geometry under 10 directional lightings from different angles. This process generates 30 images in total for each candidate.
We then project both images and text prompt to CLIP embeddings and calculate their matching scores. Specifically, given a face candidate $i$ with parameter $\bm{\beta}_i$. We render the corresponding mesh $T({\bm \beta}_i)$ using a mesh renderer $\mathcal{R}_m(\cdot)$ with the predefined 3 camera poses $c \in \mathcal{I}_m$ and 10 lightings $l \in \mathcal{J}_m$. The rendering process then is: $\mathbf{I}_i^{c,l} = \mathcal{R}_m(\bm{\beta}_i, c,l)$, where $\mathbf{I}$ denotes the rendered images.
Then we embed the images using the CLIP image encoder $\mathcal{E}_\text{vison}$ and average the generated latent codes for each candidate geometry (to guide the CLIP image encoder to focus more on geometry instead of appearances), which can be formulated as follows:
\begin{equation}
    e_{i} = \mathbb{E}_{c,l} \big [  \mathcal{E}_\text{vison}( \mathcal{R}_m(\bm{\beta}_i, c,l) ) \big ],
\end{equation}
where ${\mathbb{E}}(\cdot)$ represents the expected value computation. The text embedding then is simply obtained through encoding the user prompt as $e_\text{t} = \mathcal{E}_\text{text}(\mathcal{P})$, where $\mathcal{E}_\text{text}$ is the text encoder of CLIP. Instead of calculating correlations between $e_i$ and $e_\text{t}$, we compute in a relative way 
following AvatarClip~\cite{hone2022avatarclip} as:
\begin{equation}
    s = \lambda_d s_d+\lambda_r s_r, \ \ \text{where} \ \ 
    s_d = \tilde{e_i} \cdot \tilde{e_\text{t}},  \ \ 
    s_r = \tilde{ \Delta e_i } \cdot \tilde{ \Delta {e_\text{t}} }
\end{equation}
and $\tilde{x}$ represents the normalized value of $x$, $\Delta e_i = e_i - \bar{e_i}$, $\Delta e_\text{t} = e_\text{t} - \bar{e_\text{t}}$, and $\bar{e_i}$, $\bar{e_\text{t}}$ are the anchor embeddings, i.e., that of the anchor text, i.e., ``the face'', and the mean mesh $\mathbf{\bar T}$. We use the combined value $s$ as the final matching score.
Finally, we select the candidate with the maximum matching score as the coarse geometry $\mathbf{T}^*$, where the corresponding identity code is $\bm{\beta}^*$.

\subsection{Detail carving}
\label{subsec:detailcarving}
We then carve details on the selected coarse geometry $\mathbf{T}^*$ by optimizing additional vertex displacements and detailed normal maps in the tangent space.
Recall that the coarse mesh is selected from a set of randomly generated geometries using parametric representations. The selected coarse mesh is over-smoothing and deviates from the text prompt. To increase realism and make the geometry more closely match the input prompt, we develop a detail carving process. 
We model the face details using two components, i.e., the vertex displacement $\mathcal{V}_d$ and geometric detailed tangent space normal map $\mathcal{N}_d$, on top of the coarse mesh $\mathbf{T}^*$. 
The vertex displacement imposes geometric correction to $\mathbf{T}^*$, while the geometric detailed tangent space normal map provides more details during rendering, such as deep wrinkles. 
As illustrated in Fig.~\ref{fig:geometry}, we use the generic LDM, i.e., Stable Diffusion, to add facial details to the coarse geometry with prompt guidance. 
%
Specifically, the vertex displacement and tangent space normal map takes effect during the face rendering process, and hence gradient of the generic LDM with rendered faces as input can be passed using the Score Distillation Sampling technique for learning $\mathcal{V}_d$ and $\mathcal{N}_d$.
%
%

With the selected coarse geometry, 
our detailed geometry is formulated as:
\begin{equation}
    \mathbf{ T^\dagger}=\mathbf{T^*}+\mathcal{V}_d \odot \mathbf{n}(\mathbf{T^*}),
\end{equation}
where $\mathbf{n}(\cdot) \in \mathbb{R}^{3\times N}$ represents the vertex normal, 
$\odot$ represents element-wise multiplication. 
%
%
Then with the corrected mesh with vertex displacement added, we can render its images given a camera pose $c$ and lighting direction $l$:
\begin{equation}
\mathbf{I} = \mathcal{R}_m(\mathbf{ T^\dagger}(\mathcal{V}_d), \mathcal{N}_d, c,l),
\end{equation}
where $\mathcal{R}_m$ is the differentiable render using tangent space normal maps~\cite{Laine2020diffrast,munkberg2021nvdiffrec}.
We can write the SDS loss of generic LDM on rendered image $\mathbf{I}$ as follows:
\begin{equation}
{\nabla}_{x_d} \mathcal{L}_\text{SDS} (\mathbf{I}) \overset{\eqtr}{=} \mathbb{E}_{t, \epsilon} {\bigg{[}} w(t) \big{(} {\epsilon}_\phi(z_t^d; t,\mathcal{P})-\epsilon \big{)} \frac{\partial{\mathbf{z}_t^d}}{\partial{\mathbf{I}}} 
 \frac{\partial{\mathbf{I}}}{\partial{x_d}} {\bigg{]}}
\end{equation}
where $x_d=[\mathcal{V}_d, \mathcal{N}_d, {\bm \beta}^*]$ are the optimizing parameters, $z^d=\mathcal{E}(\mathbf{I})$ is the encoded image using LDM image encoder,
$w(t)$ is a weighting depending on discrete time step $t$, $\epsilon_\phi$ is the denoiser of generic LDM  with classifier-free guidance. 
Notice we also refine the shape parameters ${\bm \beta}^*$, denoted as ${\bm \beta}^\dagger$.
%

\begin{figure*}[t]
    \centering
    \includegraphics[width=\linewidth]{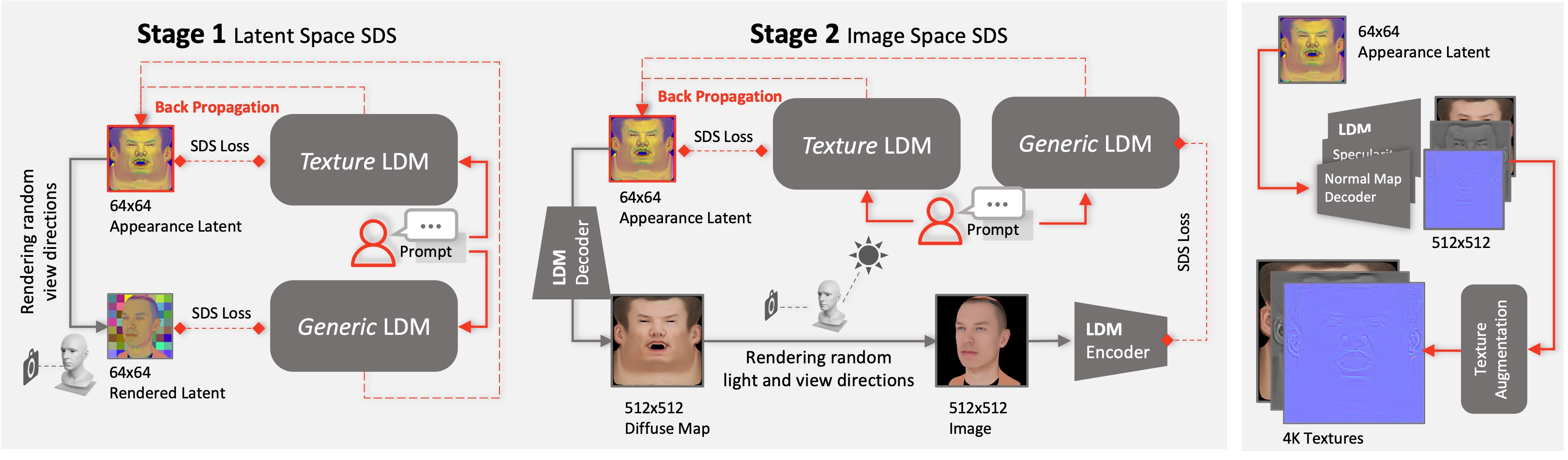}
    \caption{
The overview of physically-based texture diffusion. 
To generate detailed and realistic textures that match the input prompt, DreamFace performs Dual-path SDS on textures with the use of both a generic LDM and a texture LDM, in both the latent space and image space. By jointly optimizing using two LDMs, we are able to generate high-quality diffuse texture maps that match the input prompt and are consistent with UV unwrapping. An additional texture translation and augmentation module are also included to generate all physically-based textures with high resolution, suitable for rendering.
    }
    \label{fig:texture}
\end{figure*}

%
During the training process, for each rendering we randomly sample a camera pose $c$ and lighting direction $l$. The sampling space of the camera pose is constrained to an arc with endpoints from the left $45^\circ$ to the right $45^\circ$ w.r.t. the front-facing view.
The lighting directions $l$ are purely random within the hemisphere in the front part of the face.
Besides the SDS loss, we further add regularization terms to ensure the rationality of generated details. The additional regularization losses include:
\begin{gather}
    \mathcal{L}_{sha}=\|\bm{\beta}^\dagger-\bm{\beta}^*\|_2^2,  \,\,\,\,\,\,
    \mathcal{L}_{geo}=\text{Laplacian}(\mathbf{T}^\dagger,\mathbf{T}^*), \, \nonumber \\
    \mathcal{L}_{map}=\|\Delta \mathcal{N}_d\|_2^2+\|\nabla \mathcal{N}_d\|_2^2,
\end{gather}
where $\text{Laplacian}(\cdot)$ represents the Laplacian smooth loss between two meshes, and $\mathcal{L}_{map}$ regularizes both the gradient and divergence of the detailed normal map for smoothing. 
The final optimization objective is defined as follows:
\begin{equation}
\mathcal{L}_{carve}=\mathcal{L}_\text{SDS}+\mathcal{L}_{sha}+\mathcal{L}_{geo}+\mathcal{L}_{map},
\end{equation}
where the corresponding weights of each term are ignored for clear presentation.
Following DreamFusion~\cite{poole2022dreamfusion}, we uniformly random the discrete time step $t\sim \text{Uniform}(0.02 t_\text{max},0.98 t_\text{max})$ when optimizing.
%
The detail carving process results in the generation of a detailed geometry that closely matches the input prompt and effectively conveys the described facial characteristics. This detailed geometry is of high quality and is suitable for use in the subsequent appearance generation process.

\subsection{Hair selection}

In addition to generating the facial geometry, our pipeline also includes the generation of a realistic hairstyle that matches the input prompt. Similar to the geometry generation process, we use CLIP to select the hairstyle candidates with the highest matching score with the input prompt. Our hair dataset comprises 16 hairstyles created by professional artists. We first select the hairstyle that best matches the prompt and then select the predefined hair color by rendering the hair on the head of the geometry. This results in a detailed facial asset with a corresponding hairstyle that closely matches the input prompt.

\section{Physically-based Texture Diffusion}
\label{sec:appearance}

%
%

Appearance is a critical aspect of animatable neural facial assets, as it allows people to recognize and distinguish faces in a single glance. 
%
This section aims to generate an appearance, controlled by the diffuse, specularity and normal maps in texture space in addition to the detailed geometry, that closely matches the input prompt. Beyond effectively conveying the described characteristics of the input prompt, the consistency of texture maps with UV unwrapping is necessary to ensure compatibility with the existing computer graphics production pipeline. 
Therefore, we propose a dual-path mechanism to jointly optimize texture using two kinds of diffusion models, one generic diffusion model for diverse generation ability from general prompt inputs, and a novel texture one to ensure the textural specifications in the UV space, to predict the neutral facial assets that are consistent with both the predicted geometry and text prompt. Moreover, for efficient appearance generation, inspired by Latent-NeRF~\cite{metzer2022latent-nerf}, we further introduce a two-stage optimization to perform Score Distillation Sampling (SDS) in both the latent and image spaces, where the latent one significantly provides compact priors for fine-grained synthesis. 
In the following sections, we will first introduce how to pretrain a diffusion model in the texture space (Sec.~\ref{subsec:TextureLDM}) and then present our efficient two-stage optimization scheme with dual-path enhancement (Sec.~\ref{subsec:Twostage}). \textcolor{black}{Finally, we generate physically-based textures with high resolution in Sec.~\ref{subsec:PhyTexture}}.

\subsection{Learning diffusion model in texture space}\label{subsec:TextureLDM}

The synthesis of image from user-defined text prompt has significantly progressed via text-conditional diffusion models (DMs) ~\cite{nichol2021glide, ho2022cascaded, saharia2022imagen}, where DMs break down the generation problem into a sequential denoising process. 
Using DMs for text-conditional generation is much easier compared with other deep generative models, as DMs can be easily scaled up and trained stably on billions of image-text pairs to model complex distributions of them. Therefore, we build up our texture generator on the DM pretrained on large-scale image-text pairs.
%
Moreover, denoising at the texture resolution results in unbearable computational costs. We follow the Latent Diffusion Model (LDM) technique~\cite{rombach2022stablediffusion} to conduct the diffusion process in latent space, where an autoencoder with $\mathcal{E}$ to encode the textures into latent codes, and $\mathcal{D}$ to restore into textures from latent codes. 
%
%


However, one remaining issue is that the LDM pretrained on natural images is unaware of texture specifications that define how the parts in texture correspond to semantics on face geometry. To enforce the texture specifications and meanwhile preserve the generating ability, we first collect a diverse UV texture dataset and then train our texture LDM by finetuning the pretrained LDM on the dataset to supervise the conformity of texture specifications. 

%

\begin{figure}[t]
    \centering
    \includegraphics[width=\linewidth]{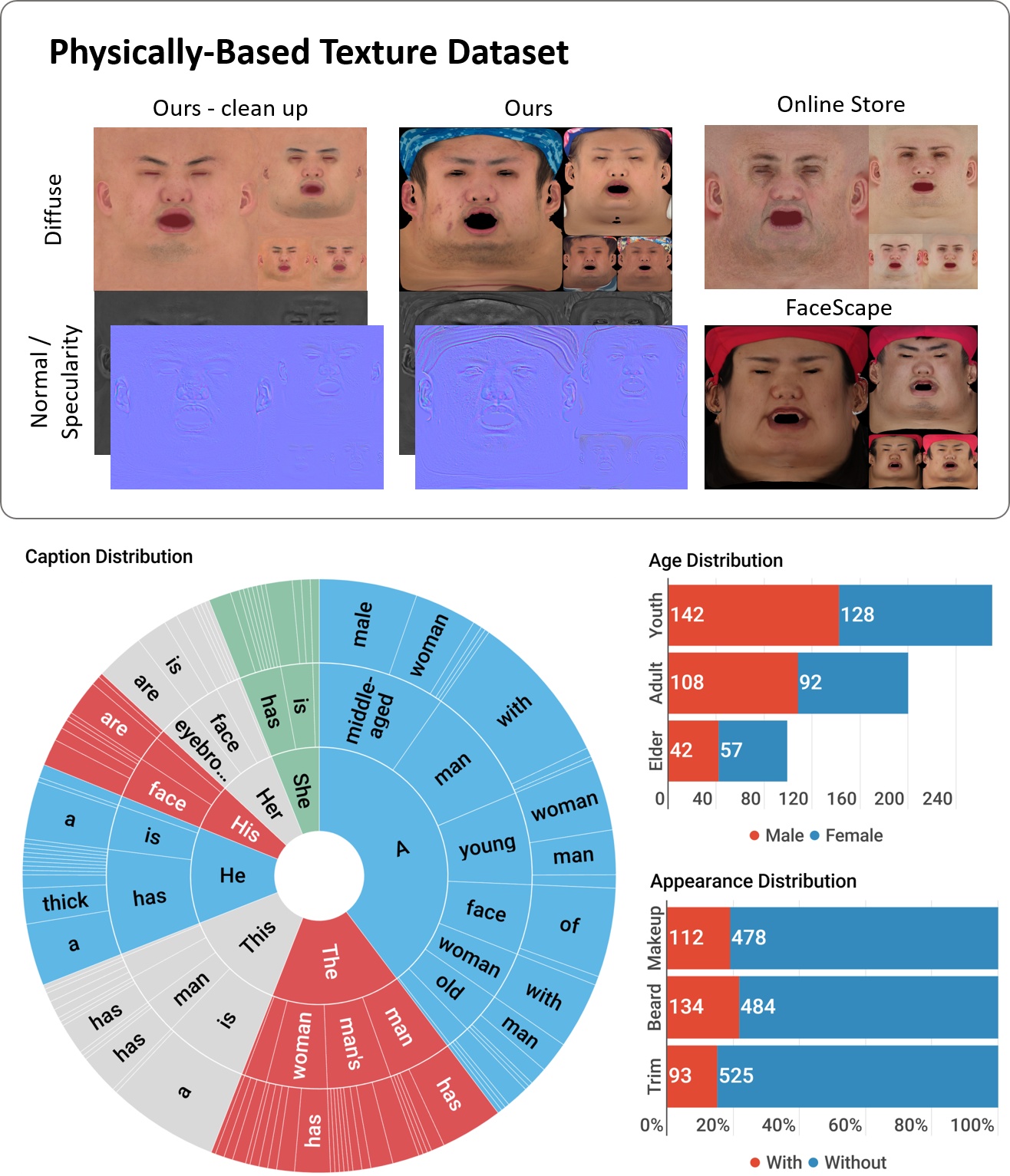}
    \caption{
Our collected physically-based texture dataset. The upper figure demonstrates several samples from several sources, which are made under different standards. The bottom left figure demonstrates the distribution of annotated text prompts. The bottom right bars illustrate the distribution of textures.
    }
    \label{fig:dataoverview}
\end{figure}

\subsubsection{Data Collection}
We collect the UV texture dataset from multiple sources, including our captured facial scans using the multi-view photometric capture system, public datasets~\cite{facescape}, and textures from commercial datasets~\cite{3dscanstore}.
By combining data from diverse datasets, the proposed dataset comprises a diverse range of skin tones, ages, and genders. However, the data merged from the other various datasets were acquired using different standards and exhibited different forms, including variations in UV unwrapping, artistic modifications, and lighting conditions. To ensure consistency and appropriateness for our research purposes, our team of expert artists and researchers performed extensive unification and annotating. A more detailed analysis of the dataset is illustrated in Fig.~\ref{fig:dataoverview}.
Our unified dataset can be formulated as $\mathcal{U} = \{ \bm{U}_\text{d}, {\bm{U}}_\text{s},{\bm{U}}_\text{n}\} $,  
where $\bm{U}_\text{d},\bm{U}_\text{s},\bm{U}_\text{n}$ represent the diffuse map, specularity map and normal map, respectively. 
Next, to collect text prompts $\mathcal{T}$ of textures, we first render the textures with geometries using our mesh and texture render and then ask annotators to generate the corresponding text descriptions following specific rules. For the textures without geometries, we render them with a template human face mesh. 

\paragraph{Non-face region masking}
In addition to texture-text pairs $(\mathcal{U}, \mathcal{T})$, we additionally collect non-face region masks ${\bm{B}}$ of textures, which include hairs, caps, markers, holes, missing edges, etc. These non-face regions in textures distract the texture LDM from learning facial appearances and make the training less stable. Hence, we combine the face color detection model~\cite{facecolordetection} with a texture component mask to accurately extract the mask of the non-face region. We condition the texture LDM on masks ${\bm{B}}$, in which the non-face regions are marked as zero and the rest as one, to let the LDM be aware of the non-face regions. We will describe this process in detail in the following section.

%

 \subsubsection{Prompt Tuning}
 \label{subsec:prompttuning}
A sizable portion of diffuse maps in our dataset containing unwanted lighting effects and deviating from strict diffuse maps is one of the biggest challenges in producing diffuse maps for use in practical applications. Many of the data in our collection were not captured using a system like Light Stage~\cite{ma2007rapidacquistion}, which employs polarized patterns to get rid of specularity and produce pure diffuse lighting. 
As a result, our dataset of diffuse maps contains a proportion of textures that exhibit undesired lighting effects and irregularities that are not suitable for our texture generation pipeline. However, we aim for the texture LDM to learn from a diverse range of textures while ensuring that generated textures during inference do not contain these unwanted artifacts.
To address this issue, we first identify two dual domains: those originating from the desired domain, where lighting and specularity have been removed, and those from the undesired domains, where lighting has not been removed, denoted as $\mathbf{\Omega}_\text{d}$ and $\mathbf{\Omega}_\text{u}$, respectively.
Then we provide a novel Prompt Tuning method to overcome this problem and guarantee that our trained texture LDM can create diffuse maps inside the desired domain $\mathbf{\Omega}_\text{d}$ only but can be trained on all the texture-text pairs in both desired and undesired domains $\{\mathbf{\Omega}_\text{d}, \mathbf{\Omega}_\text{u}\}$.

\begin{figure}[t]
    \centering
    \includegraphics[width=\linewidth]{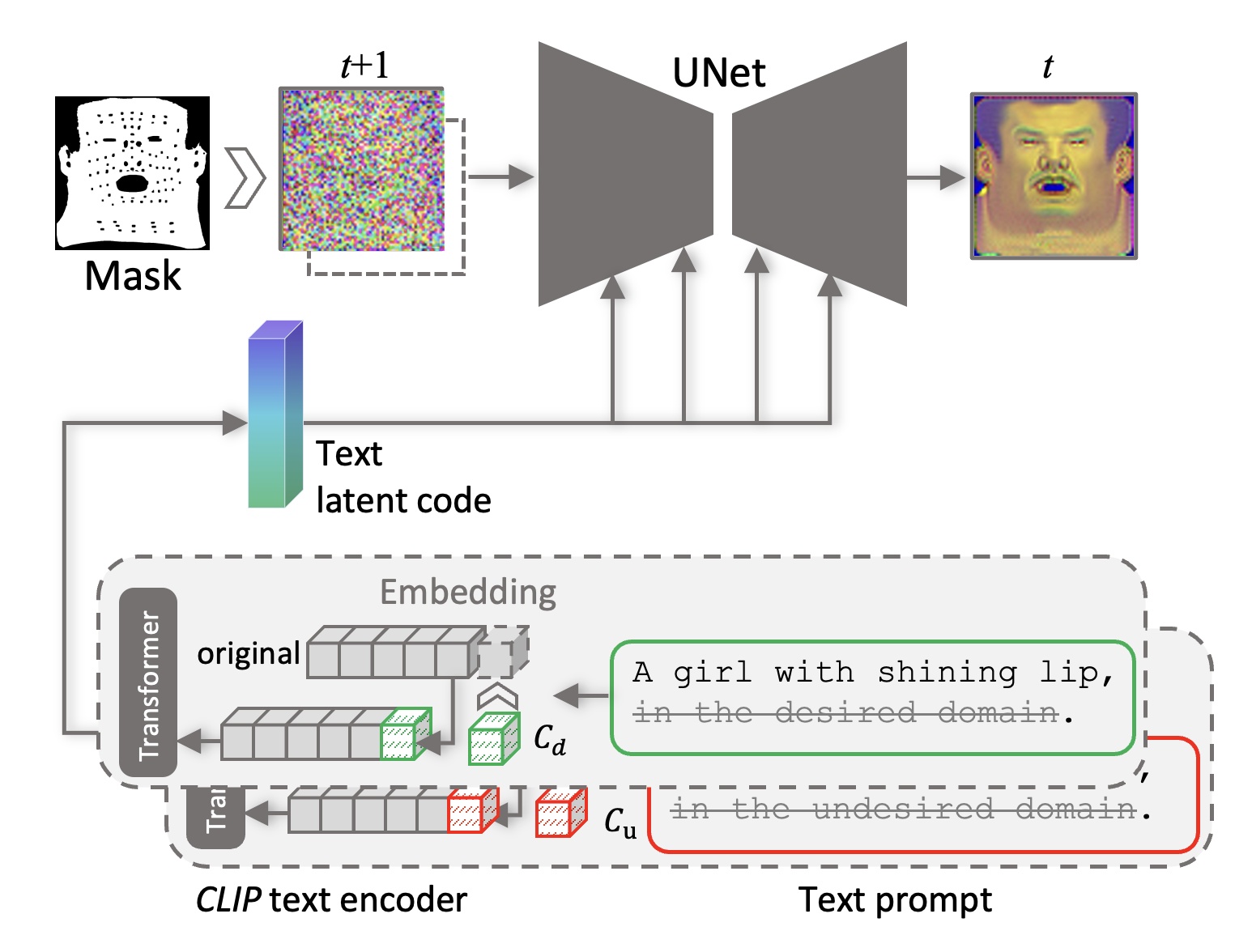}
    \caption{
    {\color{black} The overview of our Texture LDM training pipeline. Our approach utilizes two methods to generate high-quality diffuse maps: 
    (1) Prompt Tuning, instead of handcraft domain-specific text prompts, two domain-specific continuous text prompts $\mathcal{C}_\text{d}$ and $\mathcal{C}_\text{u}$ are combined with corresponding text prompt, which will be optimized during U-Net denoiser training to avoid unstable and time-consuming prompt engineering for handcraft prompt generation. 
    (2) Non-face region masking, the denoising process of LDM will be additionally conditioned on a non-face region mask to ensure that the generated diffuse map is free of any undesired elements.
    }
    }
    \label{fig:tsoverview}
\end{figure}

As shown in Fig.~\ref{fig:tsoverview}, one straightforward idea is to design two extra domain-specific prompts to indicate the textures in one particular domain explicitly, such as ``$\mathcal{T}$ in the desired domain'' and ``$\mathcal{T}$ in the undesired domain''.
However, the characteristics of the domain may not be accurately captured through the text encoder, i.e., the CLIP model here, from the handcraft prompts defined based on human understanding, which will be evaluated in Sec.~\ref{sec:evaltextureldm}. In addition, identifying appropriate handcraft prompts via prompt engineering is time-consuming and unstable, which is not practical for our needs. 
Therefore, instead of using the handcraft text prompts, we utilize prompt tuning to learn continuous text prompts, i.e., the word embedding vectors $\mathcal{C}_\text{d}$ and $\mathcal{C}_\text{u}$, to represent the two domains, respectively. Then, the domain-specific text prompts $\mathcal{C}=\{\mathcal{C}_\text{d}, \mathcal{C}_\text{u}\}$ are combined with the collected ones $\mathcal{T}$ to represent and form the domain-specific texture-prompt pairs $(\mathbf{U}_\text{d}, \mathcal{Q}(\mathcal{T}, \mathcal{C}))$, where $\mathcal{Q}(\cdot, \cdot)$ is the function that tokenizes the prompts $\mathcal{T}$ and concatenates them with the corresponding domain-specific text prompts. In particular, the domain-specific prompt $\mathcal{C}_\text{d}$ is concatenated to the prompt belonging to the $\mathbf{\Omega}_\text{d}$, and vice versa.

To learn a LDM to generate diffuse maps $\mathbf{U}_\text{d}$, the modified learning objective conditioned on prompts $\mathcal{Q}(\mathcal{T}, \mathcal{C})$ and the non-face mask ${\bm{B}}$ is formulated as follows:
\begin{equation}
\mathcal{L}_\text{LDM}(\epsilon_\theta, \mathcal{C})=\mathbb{E}_{(\mathbf{U}_\text{d},\mathcal{T}),t}\Big[\|\epsilon_\theta(\mathcal{E}(\mathbf{U}_\text{d})_t,t,\mathcal{E}_\text{text}(\mathcal{Q}(\mathcal{T}, \mathcal{C} )), \bm{B})-\epsilon\|_2^2\Big],
\end{equation}
where the domain-specific text prompts $\mathcal{C}$ and the LDM U-Net denoiser $\epsilon_\theta$ are the optimizing parameters, $\mathcal{E}(\cdot)$ is the encoder of the LDM, and $\mathcal{E}_\text{text}(\cdot)$ is the CLIP text encoder to encode the prompts $\mathcal{Q}(\mathcal{T}, \mathcal{C})$. 
To condition the LDM on the non-face mask ${\bm B}$, we scale ${\bm B}$ to the same size as the latent code and concatenate it to the noisy latent code at each time step for denoising. 

During inference, the trained texture LDM can generate diffuse maps in the $\mathbf{\Omega}_\text{d}$ by concatenating the learned $\mathcal{C}_\text{d}$ with user-defined text prompts. Similarly, it can create high-quality diffuse maps free of any undesired elements by offering a mask filled with ones.

\subsection{Two-stage Dual-path Appearance Optimization}\label{subsec:Twostage}



%

Following previous works~\cite{metzer2022latent-nerf, avrahami2022blended, linearlatentapproximationhuggingface} that explored the spatial consistencies of super-pixels in latent space, 
we conduct SDS in two-stage to optimize the texture in latent space first and then refine them in image space. This two-stage optimization scheme allows computation efficiency while retaining the high quality of texture generation.

\paragraph{Dual-path Optimization} 
Unlike previous 3D generation works ~\cite{poole2022dreamfusion,metzer2022latent-nerf, lin2022magic3d,vahdat2021score} that only apply one diffusion model to conduct SDS, we propose to combine a generic LDM and a texture LDM to generate textures in a dual-path optimization scheme, where the generic LDM retains the diverse generation ability from general input prompts, while a texture LDM aims to ensure the textural specifications in the UV space. 
%
%
Specifically, our dual-path optimization performs SDS using both the generic LDM, i.e., the Stable Diffusion ~\cite{rombach2022stablediffusion}, and our pretrained texture LDM (see Sec.~\ref{subsec:TextureLDM}) simultaneously. Although the generic LDM is capable of generating plausible textures in image space according to general text prompts, its produced textures are inconsistent with geometry, which leads the facial elements to deviate from their corresponding positions in geometry. Our dual-path optimization scheme solves this problem by additionally employing the pretrained texture LDM on the basis of the generic one to ensure the generated texture follows the UV specifications. 
%
%
%
We apply the novel dual-path optimization in both optimization stages in the latent space and the image space, respectively, enabling the generalization of textures and consistency in UV space. In the first stage, we utilize dual-path optimization to generate textures in the latent space that provides compact priors for fine-grained synthesis. Similar to the first stage, in the second stage, we also apply the dual-path optimization to enforce UV map specifications of texture and retain generalization ability, but with a detailed normal map and random lighting applied to enhance appearance details in the image space while disentangling lighting from diffuse map.

In the following sections, we introduce the design detail of our dual-path optimization scheme in latent space SDS (Sec.~\ref{subsec:latentspacesds}) and image space SDS (Sec.~\ref{subsec:imagespacesds}), respectively. 


\subsubsection{Latent space SDS}
\label{subsec:latentspacesds}

Inspired by Latent-NeRF~\cite{metzer2022latent-nerf} exploring the spatial consistency of super-pixels and performing the denoising process efficiently, we perform both texture rendering and SDS on latent space directly via our dual-path optimization scheme. 
%
Specifically, given the generic LDM denoiser $\epsilon_\phi$ and the texture LDM denoiser $\epsilon_\theta$, we perform the SDS simultaneously but operate in different modalities, including rendered latent code for the generic LDM and texture latent code for the texture LDM. 
We denote the texture latent code to be learned as $z \in \mathbb{R}^{64\times 64\times 4}$, and the rendered latent code of $z$ with geometry and camera as $z^r \in \mathbb{R}^{64\times 64\times 4}$. 
The rendered latent code $z^r$ is generated using differentiable mesh renderer $\mathcal{R}_a$ with texture using random background augmentation similar to previous methods \cite{hone2022avatarclip,khalid2022clipmesh}, which could be formulated as follows:
\begin{equation}
    z^r = \mathcal{R}_a (\mathbf{T^\dagger},z, c),
    \label{eq:render_latent}
\end{equation}
where $\mathbf{T^\dagger}$ is the detailed geometry generated from Sec.~\ref{subsec:detailcarving}, and $c$ conforms to previous defined camera distribution also in Sec.~\ref{subsec:detailcarving}. Then, the SDS loss of both LDMs could be formulated as follows:
%
\begin{equation}
\begin{aligned}
    \nabla_{z} \mathcal{L}_\text{SDS}^g &= \mathbb{E}_{t, \epsilon} {\bigg{[}} w_r (t)(\epsilon_\phi(z_t^r; t, \mathcal{P})-\epsilon) \frac{\partial z^r}{\partial{z}} {\bigg{]}},\\
    \nabla_{z} \mathcal{L}_\text{SDS}^\tau &= \mathbb{E}_{t, \epsilon} {\bigg{[}} w_\tau (t)(\epsilon_\theta(z_t;t,\mathcal{P^\prime })-\epsilon) {\bigg{]}},
    \label{eq:loss_sds_latent}
\end{aligned}
\end{equation}
where $t$ is the uniform discrete time step shared by both LDMs, $\mathcal{P}$ is the input user prompt, and $\mathcal{P^\prime}$ is the augmented prompt for texture LDM with specifically designed keyword appended, as discussed in Sec.~\ref{subsec:prompttuning}. We compute ${\partial z^r} / {\partial z}$ from the differential renderer $\mathcal{R}_a$ in Eqn.~\ref{eq:render_latent}. 
The final optimization objective then is the combination of the two SDS losses as follows:
\begin{equation}
\mathcal{L}_\text{tex}=\lambda_g \mathcal{L}_\text{SDS}^g+\lambda_\tau \mathcal{L}_\text{SDS}^\tau,
\end{equation}
where $ \lambda_g $ and $\lambda_\tau$ are corresponding weights that require deliberate design which we will further discuss in Sec.~\ref{subsucsec:abla} and supplementary video.
We observe that following DreamFusion~\cite{poole2022dreamfusion} to uniformly sample the time step and pass the gradient leads to fragility in optimization.
In contrast, we vary the $t$ as in the denoising process, i.e., we decrease $t$ with equal intervals from $t_\text{max}$ to $0$ as in a real denoising process.
%
Finally, by performing SDS under our dual-path optimization scheme, we obtain a learned texture latent code $z$ that matches the user-defined prompt and conforms to the texture specification.


\subsubsection{Image space SDS}
\label{subsec:imagespacesds}

\textcolor{black}{Our proposed dual-path optimization scheme based on a pair of generic and texture LDMs allows efficient learning of well-formed texture latent code in the first stage. In the second stage, we apply the dual-path optimization scheme in the image space to further obtain a more detailed diffuse map. Specifically, we perform SDS of the generic LDM path on rendered RGB images and utilize the LDM autoencoder $\mathcal{E}$ and $\mathcal{D}$ to map between the latent space and the image space, while keeping the SDS process for the texture LDM path the same as in the first stage.}
%
%
%
The optimization scheme in the image space SDS follows the basic framework of that in the latent space SDS process, \textcolor{black}{but we apply detailed normal maps and random lighting to enhance appearance details while disentangling lighting from diffuse maps during general LDM SDS.}
%
%
Specifically, instead of rendering the latent code $z$ directly into $z^r$, we convert the 64$\times$64 texture latent code $z$ into a 512$\times$512 diffuse texture map using the LDM decoder $\mathcal{D}$.
%
%
Then, we render the geometry $\mathbf{T^\dagger}$ using the decoded texture into an 512$\times$512 image under random camera pose $c$ and lighting $l$ with detailed normal map applied, 
%
and the rendering process is defined as follows:
\begin{equation}
    \mathbf{I}=\mathcal{R}_a (\mathbf{T^ \dagger}, \mathcal{N}_d, \mathcal{D} (z), c,l).
\end{equation}
We then use the LDM encoder $\mathcal{E}$ to encode the rendered image $\mathbf{I}$ into latent code $z^{Hr} = \mathcal{E}(\mathbf{I})$ and perform the SDS process of the generic LDM. In particular, SDS loss of the generic LDM is
\begin{equation}
    \nabla_{z} \mathcal{L}_\text{SDS}^g = \mathbb{E}_{t, \epsilon} {\bigg{[}} w_r (t)(\epsilon_\phi(z_t^{Hr}; t, \mathcal{P})-\epsilon) 
    \frac{\partial z^{Hr}}{\partial \mathbf{I}} \frac{\partial \mathbf{I}}{\partial \mathcal{D}(z)} 
    \frac{\partial\mathcal{D}(z)}{\partial z} 
     {\bigg{]}} \\
    \label{eq:loss_sds_image}
\end{equation}
according to the Leibniz rule. 
And the SDS process of the texture SDS path is the same with the Eq.~\ref{eq:loss_sds_latent}. 
After the dual-path optimization in image space, we are able to obtain a finetuned texture latent code $z^H$ that not only has a higher level of details but also conforms to our texture space prior. Finally, the $z^H$ can be decoded into the high-quality diffuse map via the decoder $\mathcal{D}$.

\subsection{Physically-based textures generation}\label{subsec:PhyTexture}

Beyond high-quality texture optimized in Sec.~\ref{subsec:Twostage}, an indispensable component of high-fidelity facial assets is the physically-based textures including high-resolution diffuse, specularity and normal maps, which enable photo-realistic rendering using existing CG production pipeline.
There is a strong correlation between components of physically-based textures as proposed by ~\citet{li2020learningformation}. Hence we can infer the specularity and normal maps from the diffuse texture with an image-to-image translation technique, and we re-use the latent space of the LDM autoencoder for its compactness and efficiency. 

\paragraph{Texture Translation} Recall that in the dual path optimization process, we obtain a fine-tuned texture latent code $z^H$, which could be decoded to diffuse map through $\mathcal{D}$.
To further obtain the specularity and normal maps, we develop an image-to-image translation technique inspired by the previous works \cite{li2020Dynamic}. Instead of training an image-to-image translation module from scratch, we manage to utilize the compactness of the existing LDM autoencoder latent space.
Besides the pretrained LDM encoder $\mathcal{E}$ and decoder $\mathcal{D}$, we train another two specific decoders $\mathcal{D}_\text{s},\mathcal{D}_\text{n}$ that decode the existing texture latent code $z^H$ into specularity map and normal map, respectively. Both decoders are trained using our physically-based texture dataset, consisting of 370 tuples of diffuse maps $\mathbf{U}_\text{d}$, specularity maps $\mathbf{U}_s$, and normal maps $\mathbf{U}_n$, collected by the photometric multi-stereo capture system. We then encode diffuse maps into latent codes using $\mathcal{E}$ as $u_\text{d} = \mathcal{E} (\mathbf{U}_\text{d})$. And the corresponding learning objective is formulated as follows:
\begin{equation}
\begin{aligned}
\mathcal{L}_{\text{tex}} = 
	\mathlarger {\ell} (\mathcal{D}_\text{s} (u_\text{d}), \mathbf{U}_\text{s}) + 
	\mathlarger {\ell} (\mathcal{D}_\text{n}(u_\text{d}), \mathbf{U}_\text{n}), 
\end{aligned}
\end{equation}
where $\mathlarger {\ell} (\cdot)$ represents the same loss terms described in Stable Diffusion~\cite{rombach2022stablediffusion}.
%

\paragraph{Texture Augmentation} After generating specularity and normal maps from corresponding diffuse map input, we further enhance their quality and upscale them to $4$K resolution to add pore-level details while preserving the identity information.
Specifically, we first fine-tune the face restoration network RestoreFormer~\cite{wang2022restoreformer} on our dataset at $512\times 512$ resolution to enhance the facial details. Then we refine the super-resolution model Real-ESRGAN~\cite{realesrgan} using our high-resolution textures to further generate $4096 \times 4096$ physically-based textures, which are essential for photo-realistic rendering. The final normal texture synthesizes pore-level details on the face and is also represented in tangent space. We integrate it with the geometric detailed tangent space normal map $\mathcal{N}_d$ during the rendering process.

\begin{figure}[t]
    \centering
    \includegraphics[width=\linewidth]{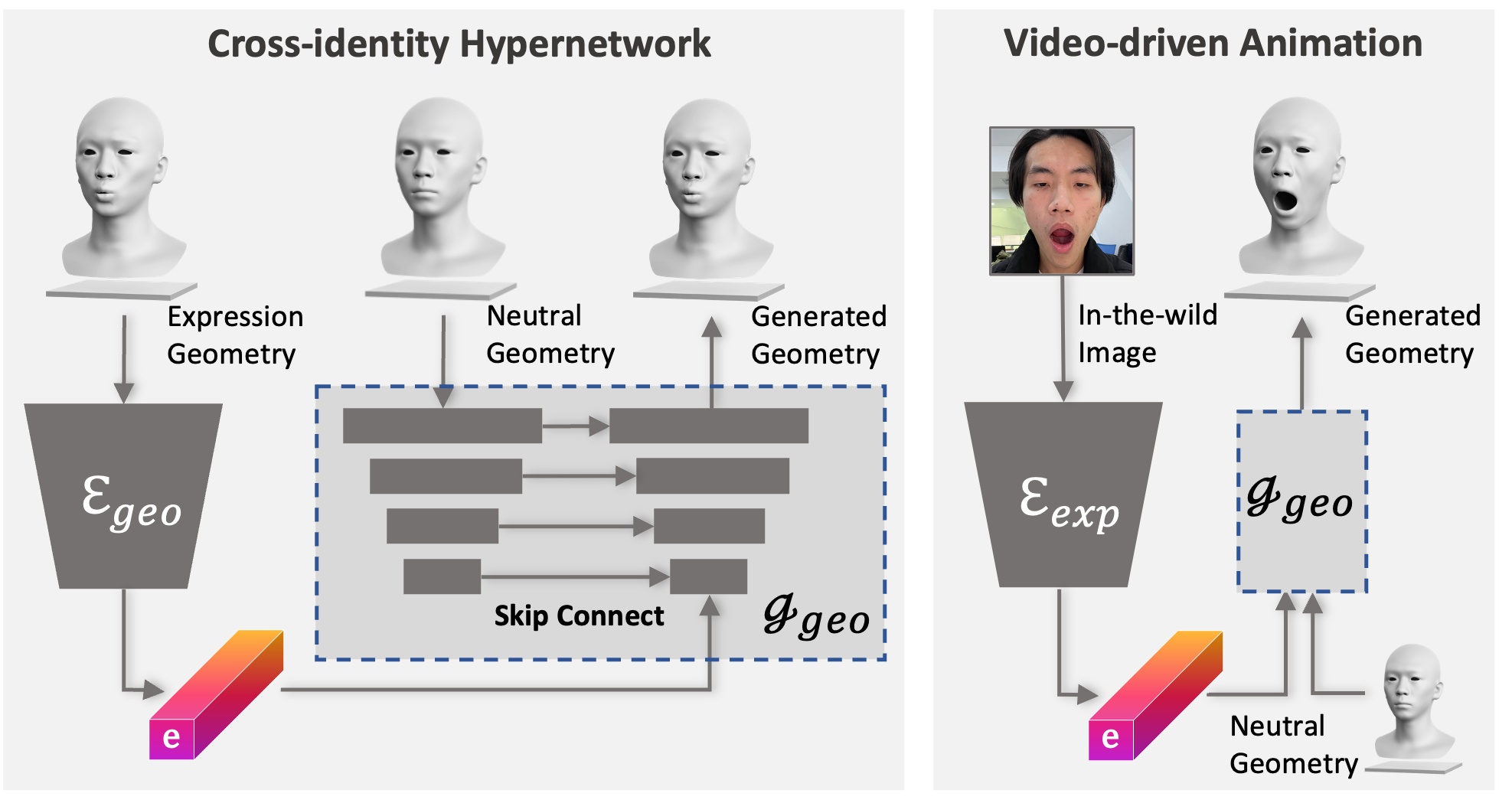}
    \caption{
Overview of animatability empowerment. First, we train a geometry generator to learn a latent space of expression where the decoder is extended to condition on neutral geometry. Then, an expression encoder is further trained to extract expression features from RGB images. Thus, we are able to generate personalized animations by conditioning on the given neutral geometry using monocular RGB images.
    }
    \label{fig:animation}
\end{figure}

\begin{figure*}[t]
    \centering
    \includegraphics[width=\linewidth]{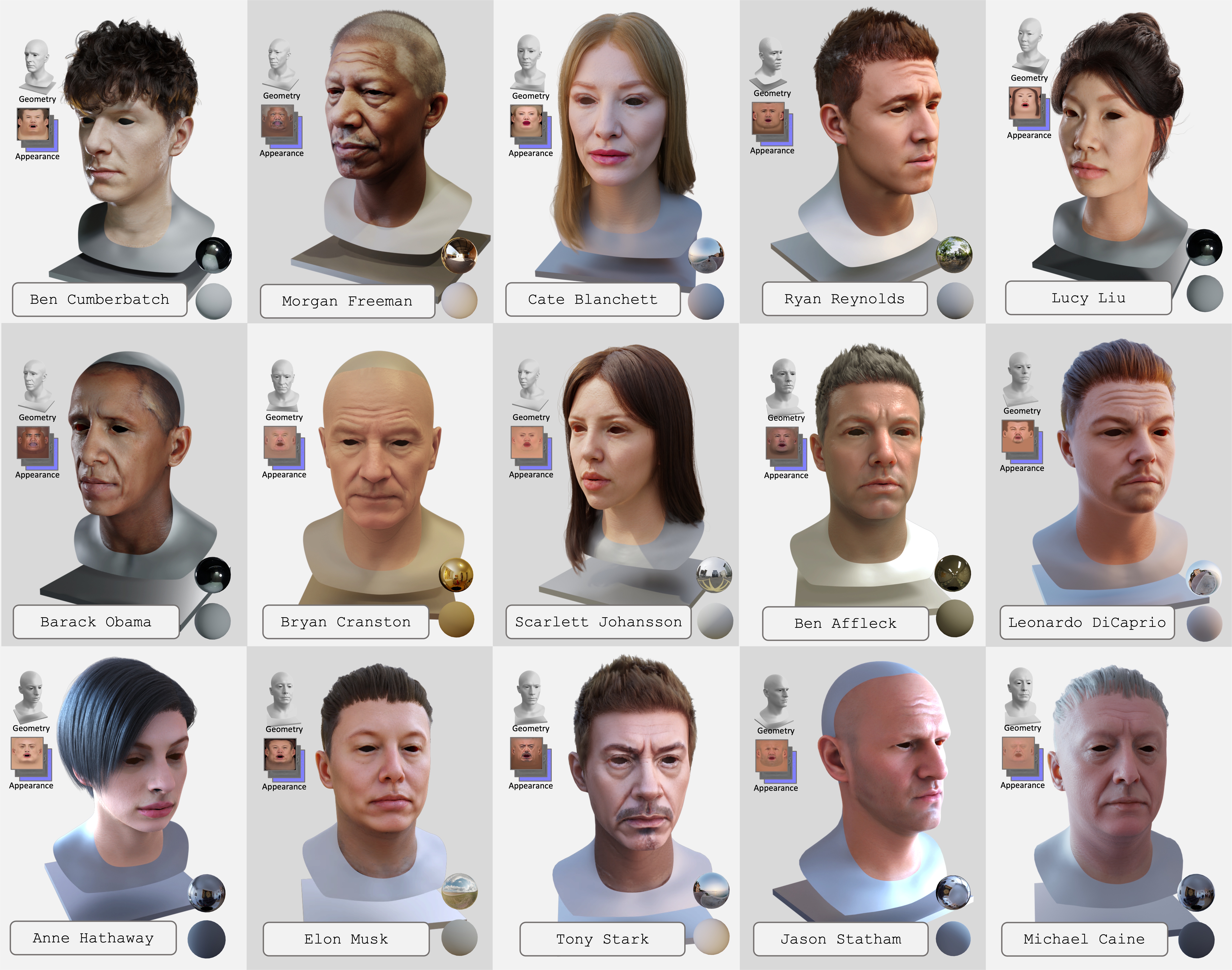}
    \caption{
    Generated facial assets of celebrities. 
    Our approach generates facial assets of celebrities that capture their personalized characteristics and achieve a high degree of resemblance. By generating physically-based textures, our facial assets achieve photo-realistic results using the modern CG rendering pipeline.
    }
    \label{fig:gallery}
\end{figure*}

\begin{figure*}[t]
    \centering
    \includegraphics[width=\linewidth]{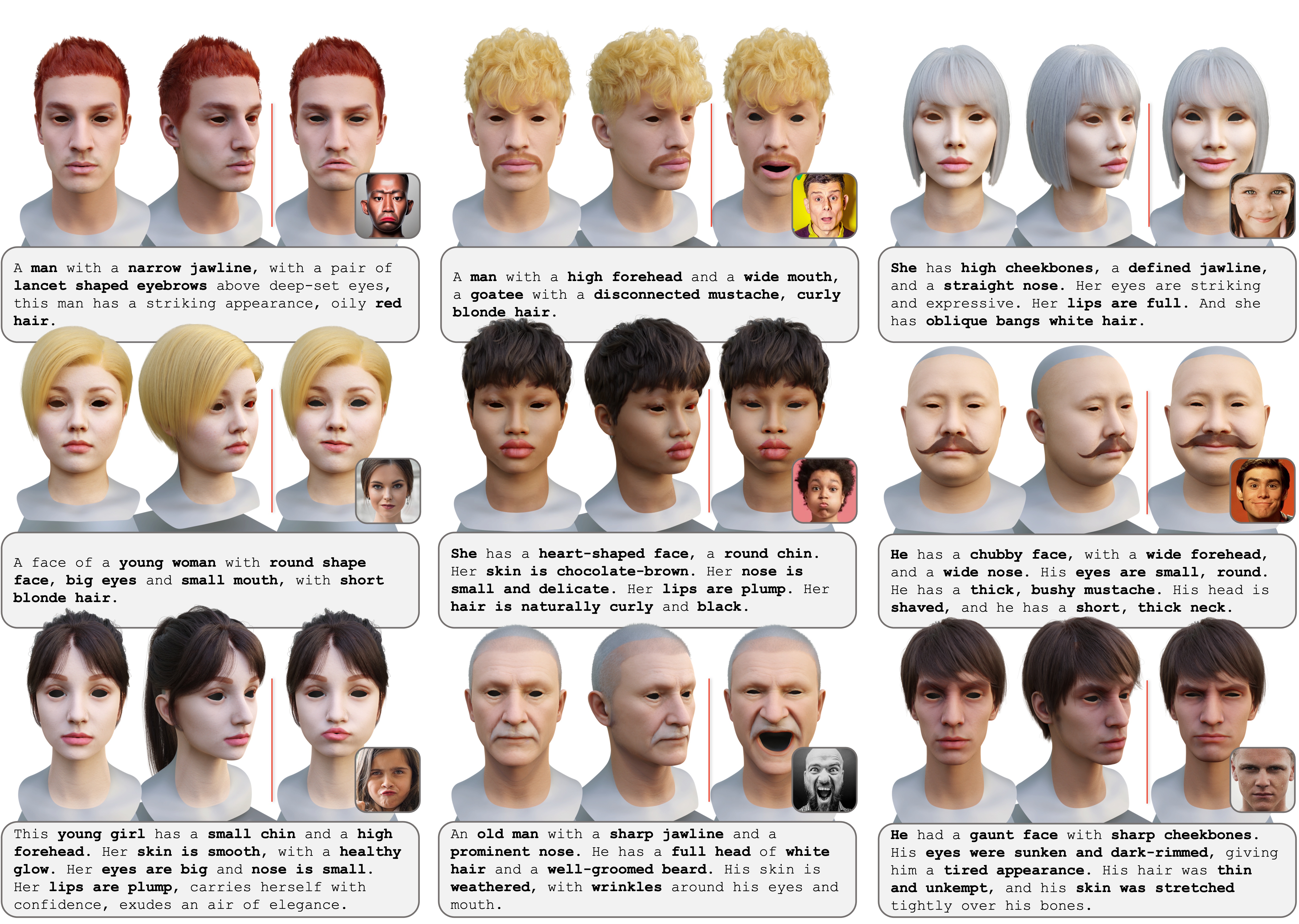}
    \caption{
    Generated facial assets from descriptions. 
    Our approach generates facial assets that faithfully match the characteristics described in the prompts.
    Through our animatability empowerment, the generated facial assets can be animated using a single RGB image and rendered photo-realistically in modern CG pipelines.
    }
    \label{fig:gallery2}
\end{figure*}

\begin{figure*}[t]
    \centering
    \includegraphics[width=\linewidth]{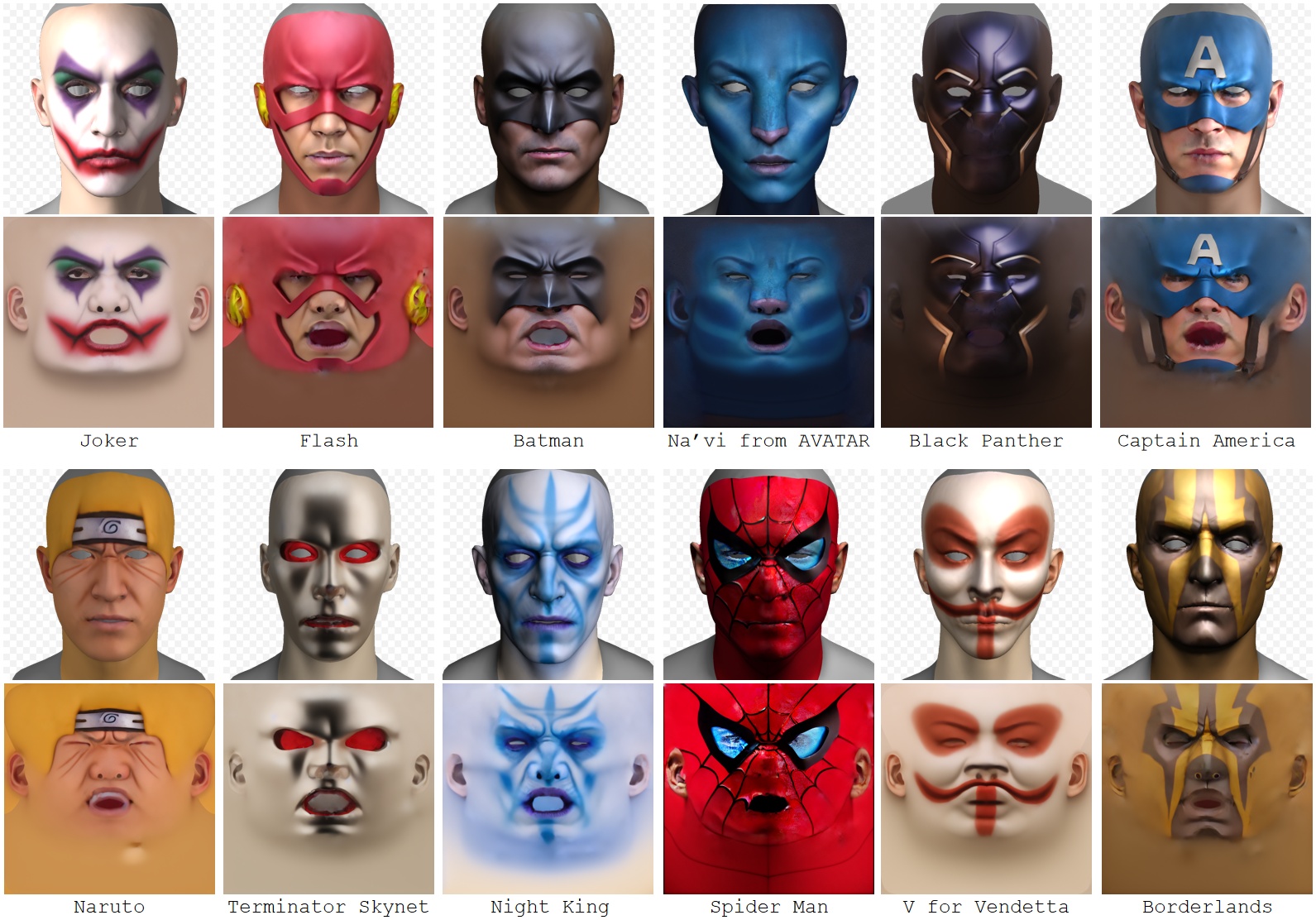}
    \caption{
Generation out of distribution.
The upper row shows the rendering results from the differentiable renderer, and the lower row shows the corresponding diffuse maps. 
Our framework faithfully reveals the facial characteristics of characters, even if they are not present in our texture dataset, for example, the pink nose of \textit{Na'vi} and the metallic patterns of \textit{Black Panther}'s mask.
In addition, our texture LDM serves as a robust prior, ensuring that the generated facial components share a consistent UV space.
    }
    \label{fig:outof}
\end{figure*}

\section{Animatability Empowerment}
\label{sec:animation}


In addition to generating fine face geometry and physically-based textures, our framework also empowers the animatability of the generated facial asset.  
While our asset directly supports traditional blendshape-based animation due to its consistent geometric topology, one could utilize existing facial performance capture techniques ~\cite{li2017flame, moser2021semi, arkit} to obtain the corresponding expression blendshape parameters from images and videos to animate our generated facial asset.

%

%
%
Beyond generic animatability using blendshapes, we explore introducing an enhanced animation scheme to produce personalized expressions while preserving the unique characteristics of our generated facial asset and creating more realistic and believable animations.



In contrast with \cite{zhang2022npfa, lombardi2018dam, laine2017di4d} which learn person-specific facial animations from extensive facial captures for one identity, we develop a cross-identity geometry hypernetwork that generates person-specific expressions and serves as a universal prior for expression space. Inspired by \citet{phonescan}, we condition the geometry generator on neutral geometry and train the hypernetwork using numerous geometries with various expressions.

%
%
%

\paragraph{Cross-identity hypernetwork}
Our geometry hypernetwork consists of a geometry expression encoder $\mathcal{E}_\text{geo}$ and a geometry generator $\mathcal{G}_\text{geo}$, where the geometry generator is basically a U-Net conditioned on neutral geometry without expressions, as exhibited in Fig.~\ref{fig:animation}.
The geometry expression encoder encodes different facial expressions into a unified expression latent code, while the geometry generator uses this code, along with the neutral geometry of a specific identity, to generate the desired facial geometry with corresponding expressions.
The forward process could be formulated as follows:
\begin{gather}
      z_e =\mathcal{E}_\text{geo}(\mathbf{G}),\,\,\,\,\,\,  \mathbf{\tilde G}=\mathcal{G}_\text{geo}(\mathbf{G}_0, z_e), \nonumber \\
    \mathcal{L}_\text{recon}=\|\mathbf{\tilde G}-\mathbf{G}\|_2^2,  
\end{gather}
where $\mathbf{G}$ is the input geometry, $z_e$ is the expression code, $\mathbf{G}_0$ is the neutral geometry of $\mathbf{G}$, $\mathbf{\tilde G}$ is the generated geometry, and $\mathcal{L}_\text{recon}$ is the difference between the generated geometry and original geometry corresponding to $z_e$ as training loss.
We train this geometry hypernetwork in a self-supervised manner, using a dataset of geometries with various expressions and identities. Once trained, the geometry generator $\mathcal{G}_\text{geo}$ is capable of producing geometries with the desired expressions for a specific identity using the unified expression latent code.

\paragraph{Video-driven animation}
With the geometry generator $\mathcal{G}_\text{geo}$ well trained, we additionally train an image expression encoder $\mathcal{E}_\text{exp}$ to extract unified expression latent code from RGB images. We freeze the geometry generator in this process, and train $\mathcal{E}_\text{exp}$ under the supervision of both real images and randomly rendered images from geometries $\mathbf{G}$ and textures $\mathbf{A}$ in the dataset, which could be formulated as:
\begin{gather}
    {\hat{z}_e} = \mathcal{E}_\text{exp}(\mathcal{R}_a (\mathbf{G, A})), \,\,\,\,\,\, 
    \mathbf{\hat G} = \mathcal{G}_\text{geo}(\mathbf{G}_\mathbf{0}, {\hat{z}_e}), \nonumber \\
    \mathcal{L}_\text{exp} =\|\mathbf{\hat G}-\mathbf{G}\|_2^2, 
\end{gather}

where $\mathcal{R}_a$ is a mesh and texture renderer that renders under random augmentations, and $\mathcal{L}_\text{exp}$ is the difference between ground-truth geometry with expression and the generated geometries. After training, $\mathcal{E}_\text{exp}$ is able to extract expression latent code from in-the-wild videos, and $\mathcal{G}_\text{geo}$ could further produce personalized animations for our generated facial asset.
%

\paragraph{Dataset}
In order to train our geometry generator with both generalizability to various identities and accurate capturing of fine-grained expression details, we capture a dataset that includes a large number of static scans of different expressions and identities, as well as several dynamic performance sequences from various performers. Our dataset consists of 38400 registered mesh and 614400 images from 300 identities across different genders, ages, and ethnicities.

\paragraph{Network training}
The geometry encoder $\mathcal{E}_\text{geo}$ consists of 7 sequential convolution layers with kernel size of 5 and stride of 2 that has [32,64,128,256,512,512,512] intermediate channels and a linear layer that maps to a latent space with 256 dimensions. The geometry generator $\mathcal{G}_\text{geo}$ is implemented as a U-Net with similar architecture as proposed by \citet{phonescan}, which down-samples input 5 times with [32,64,128,256,256] intermediate channels. The image encoder $\mathcal{E}_\text{exp}$ shares the same architecture with $\mathcal{E}_\text{geo}$.
The input and output geometry are formulated as geometry map of size 256$\times$256, as described in previous works~\cite{li2020Dynamic, li2020learningformation, zhang2022npfa}. The input image is cropped to the facial region and down-sampled to 256$\times$256.
The first stage of training the geometry hypernetwork $\mathcal{E}_\text{geo},\mathcal{D}_\text{geo}$  takes 5 days to converge. And the second stage training the image encoder $\mathcal{E}_\text{exp}$ takes 48 hours to converge. All training is done using Pytorch with AdaBelief optimizer and learning rate 5e-5, using a single Nvidia A6000 GPU.

\begin{figure*}[t]
    \centering
    \includegraphics[width=\linewidth]{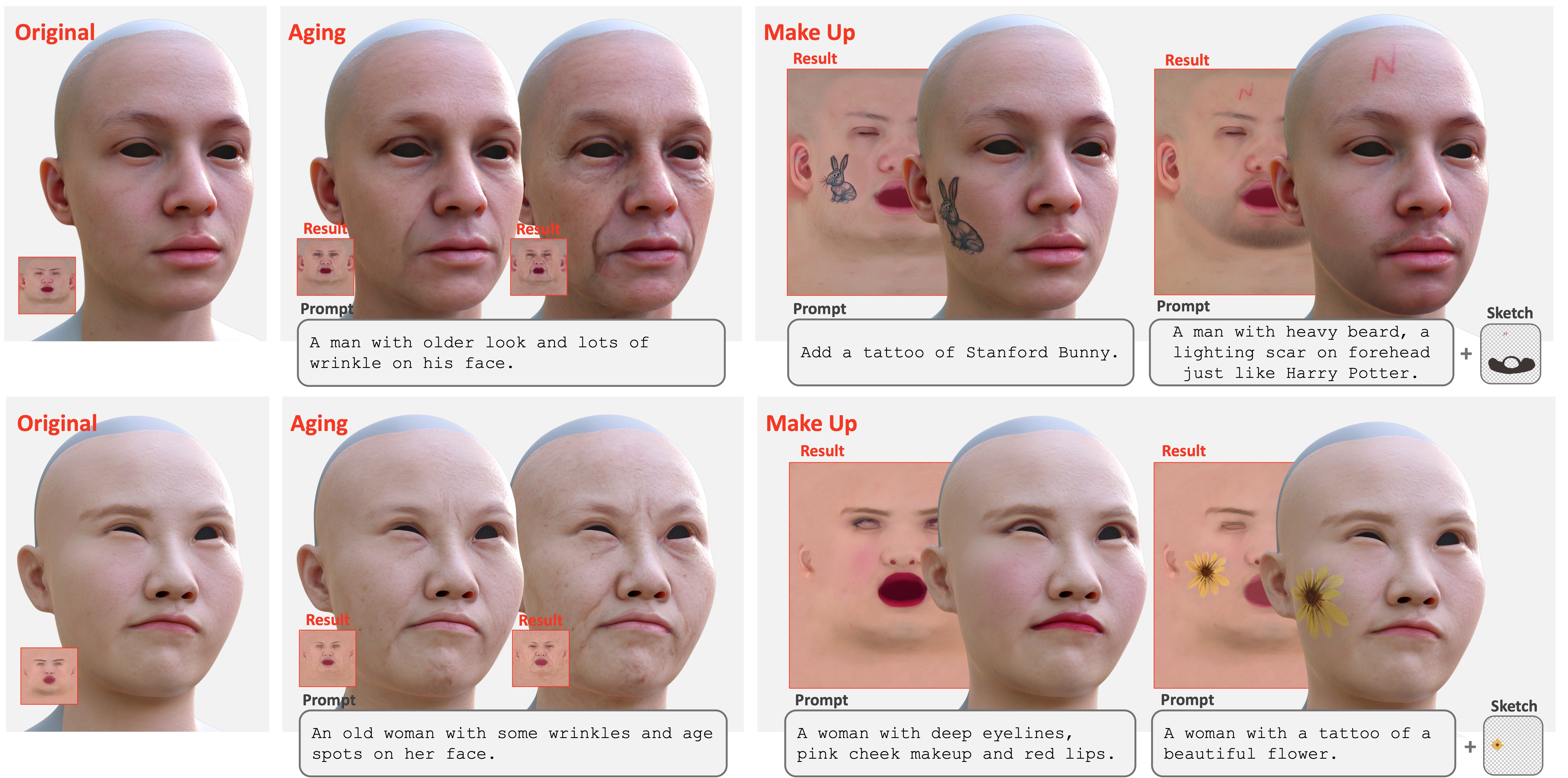}
    \caption{
Texture editing using prompts and sketches. By directly using our trained texture LDM with a prompt, one can achieve global editing effects such as aging and makeup. By further combining masks or sketches, one can create various effects such as tattoos, beards, and birthmarks. 
    }
    \label{fig:editing}
\end{figure*}

\begin{figure*}[t]
    \centering
    \includegraphics[width=\linewidth]{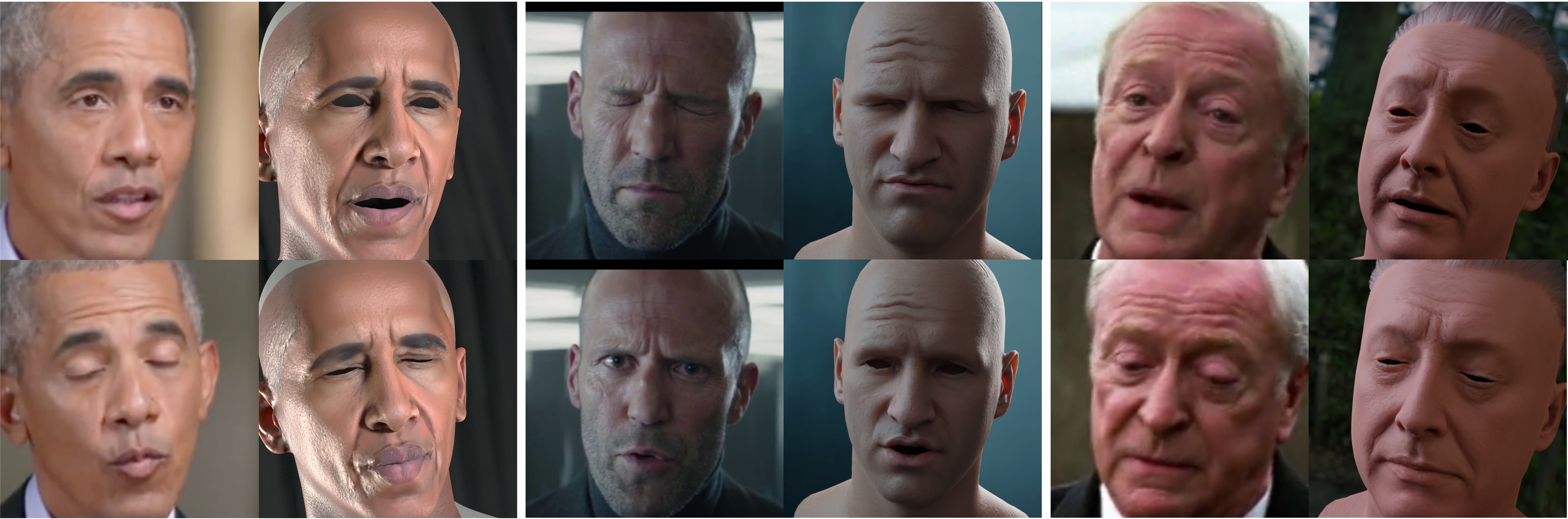}
    \caption{
Video-driven animation of generated facial assets. For each case, we show the input RGB images (left) and the personalized driven results (right). Our framework provides each generated facial asset with personalized expressions from a single image.
    }
    \label{fig:drive}
\end{figure*}

\begin{figure}[t]
    \centering
    \includegraphics[width=\linewidth]{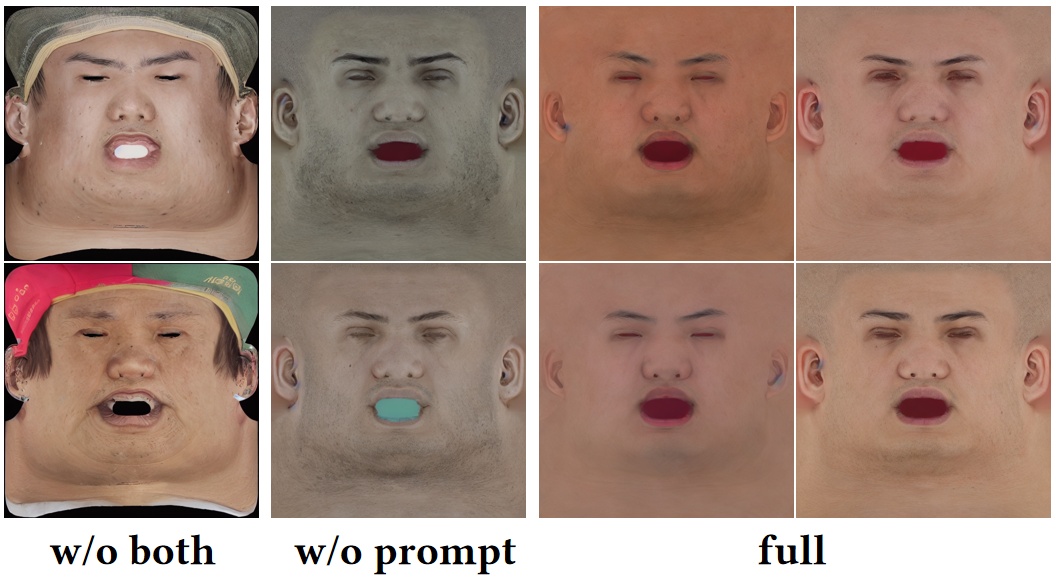}
    
    \vspace{0.5cm}
    
    \begin{tabular}{ccc}\hline\hline
    \multirow{ 2}{*}{\textbf{Training setting}}&\multicolumn{2}{c}{\textbf{KID}$\downarrow$}\\
     & {Full dataset} & {Desired domain} \\\hline\hline
       \textbf{w/o m\&p}    & 0.1282    & 0.2737 \\\hline
       \textbf{w/o prompt}  & 0.2467    & 0.1068 \\\hline
       \textbf{full}        & 0.2125    & \textbf{0.0578} \\\hline
    \end{tabular}
    \caption{
    Qualitative and quantitative comparison with different variants on learning texture LDM.
    With our two careful designs, the texture LDM produces textures with the best quality, where the lighting and unwanted elements are well removed, as illustrated in the upper figure, and also achieves the lowest KID in desired texture domain as shown in the lower table.
    }
    \label{fig:ablation}
\end{figure}

\begin{figure}[t]
    \centering
    \includegraphics[width=\linewidth]{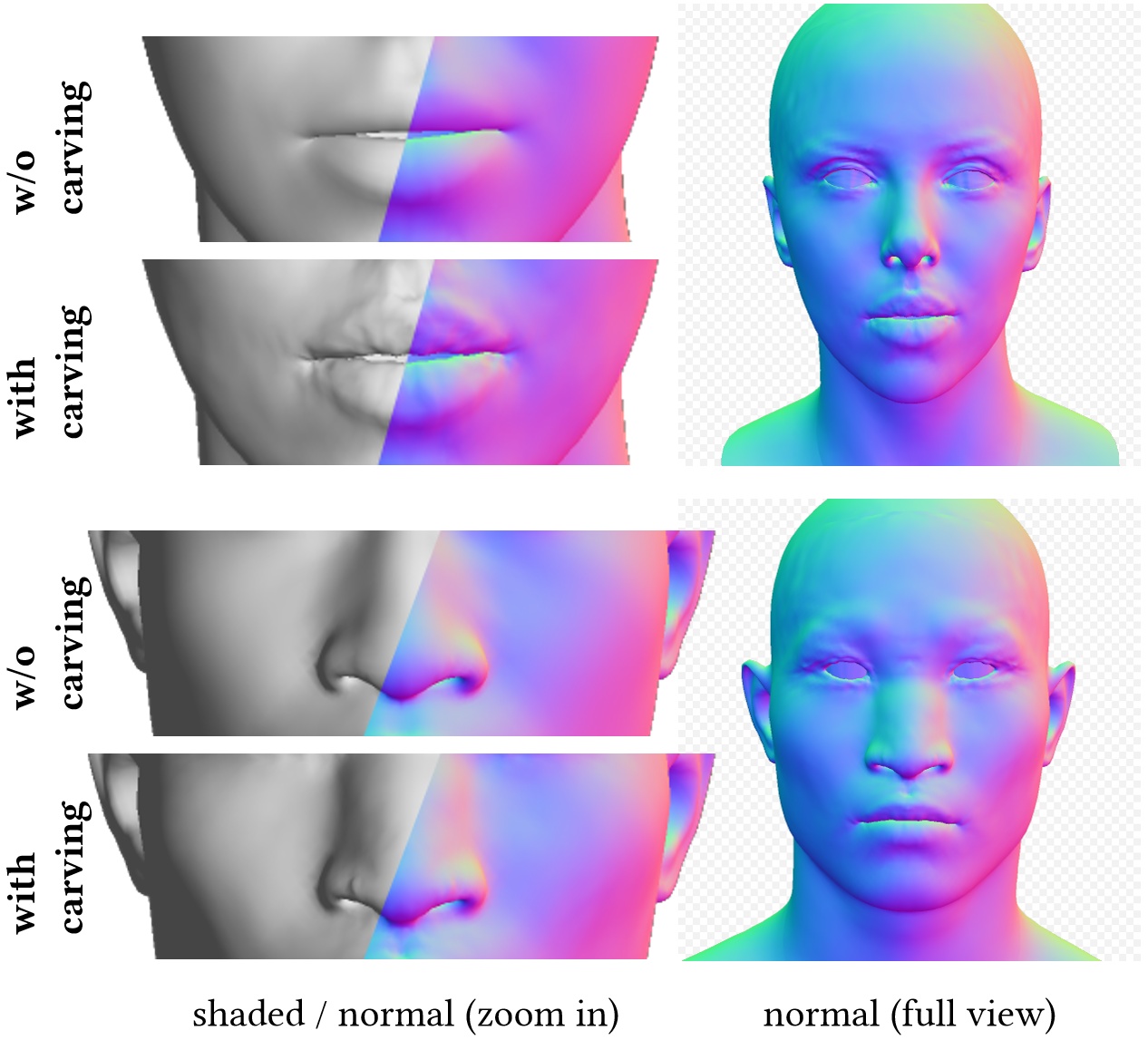}
    \caption{
Qualitative comparison on detail carving. 
For each case, we compare coarse geometry and detailed geometry with a zoomed-in view (left) and a full view (right). 
The upper case uses the prompt "the face of Scarlett Johansson" whose mouth is well-carved, demonstrating her characteristic traits. 
The lower case uses the prompt "the face with a Na'vi style nose in the movie Avatar" where a wide nose bridge resembles the indigenous alien species.
This comparison shows the effectiveness of our detail carving in capturing specific traits and features of different characters.
    }
    \label{fig:carving}
\end{figure}

\begin{figure}[t]
    \centering
    \includegraphics[width=\linewidth]{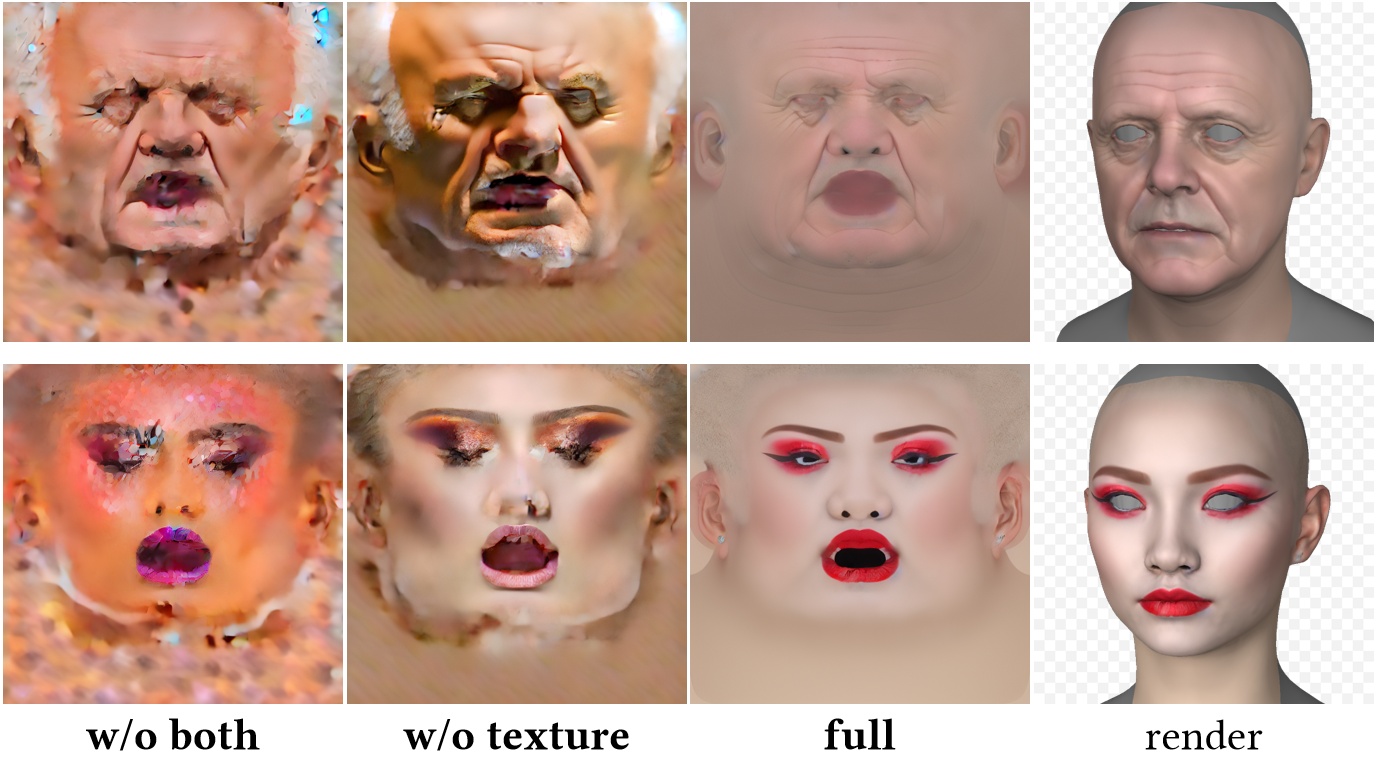}
    \caption{
Qualitative comparison on appearance latent code generation. 
For each case, we compare two variations with our pipeline together with rendering results using our differentiable renderer. The upper case has the prompt "the face of Anthony Hopkins" and the lower case has the prompt "the face of a young lady with exquisite makeup". Compared to variations without SDS in latent space and using texture LDM, our pipeline produces the best diffuse maps with precise, well-defined components in consistent UV space.
    }
    \label{fig:latent}
\end{figure}

\begin{figure}[t]
    \centering
    \includegraphics[width=\linewidth]{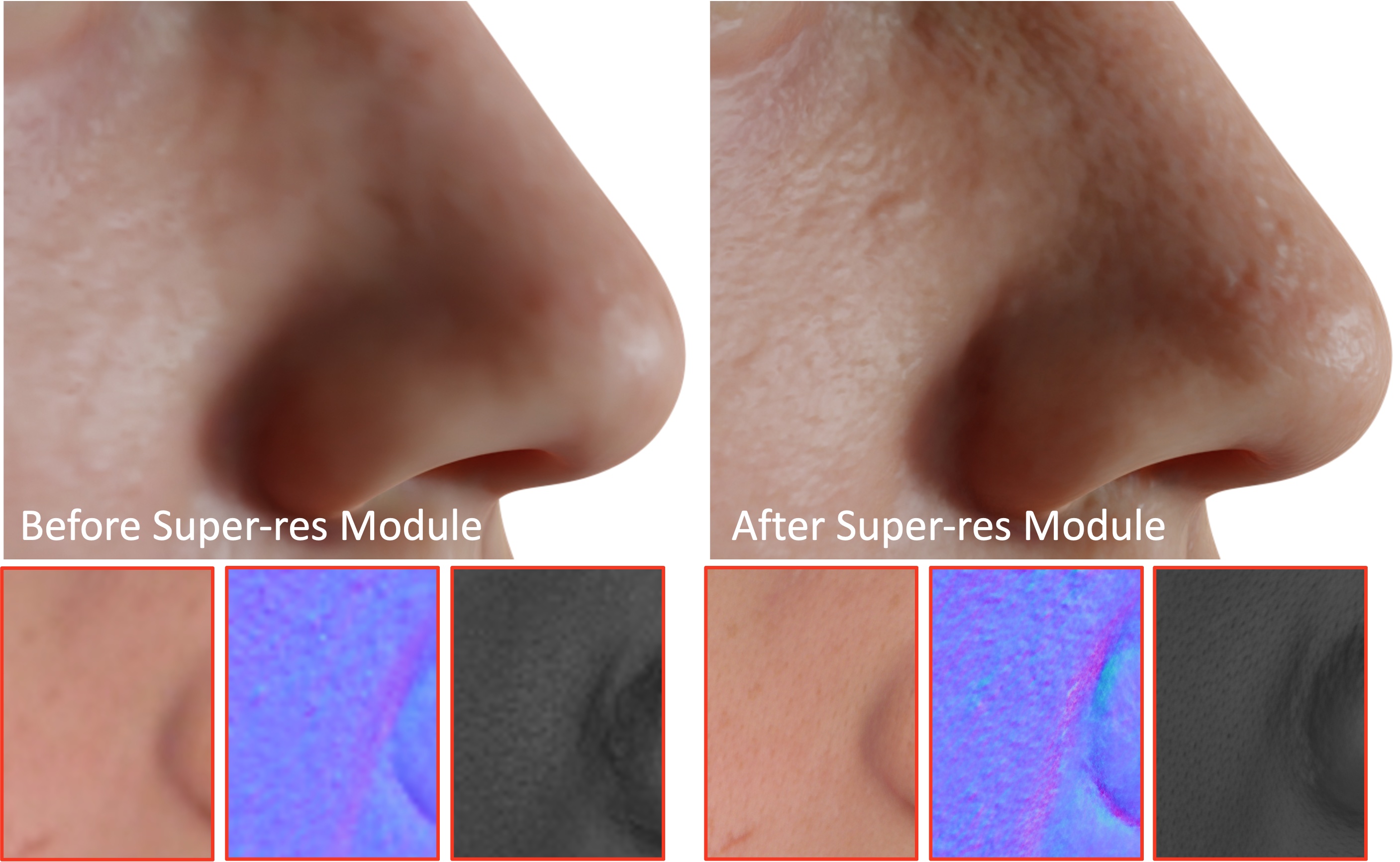}
    \caption{
Qualitative comparison of physically-based texture augmentation. 
We show the directly decoded textures from generated texture latent code (left) and the high-resolution physically-based textures (right).
    }
    \label{fig:superres}
\end{figure}

\begin{figure}[t]
    \centering
    \includegraphics[width=\linewidth]{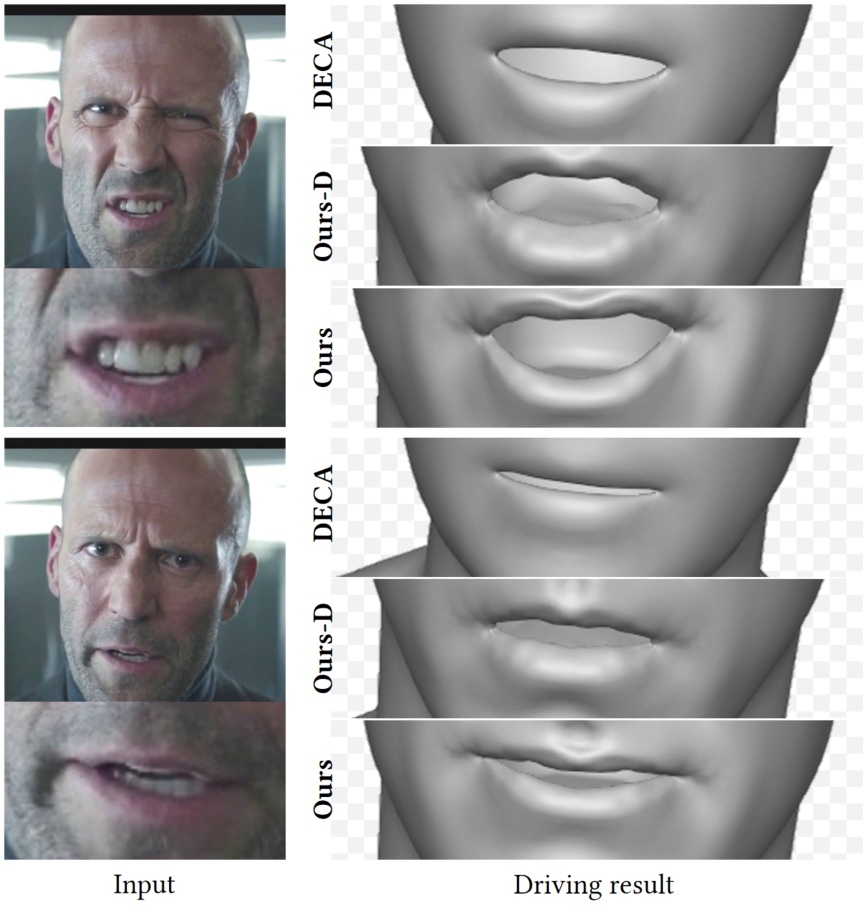}    
    \caption{
Qualitative comparison of expression generation with the previous generalized method. Compared to DECA, which uses generic blendshapes for expression control, our framework provides fine-grained expression details while carefully capturing nuanced performance, such as ``left sneering'' (upper).
    }
    \label{fig:deca}
\end{figure}

\begin{figure}[t]
    \centering
    \includegraphics[width=\linewidth]{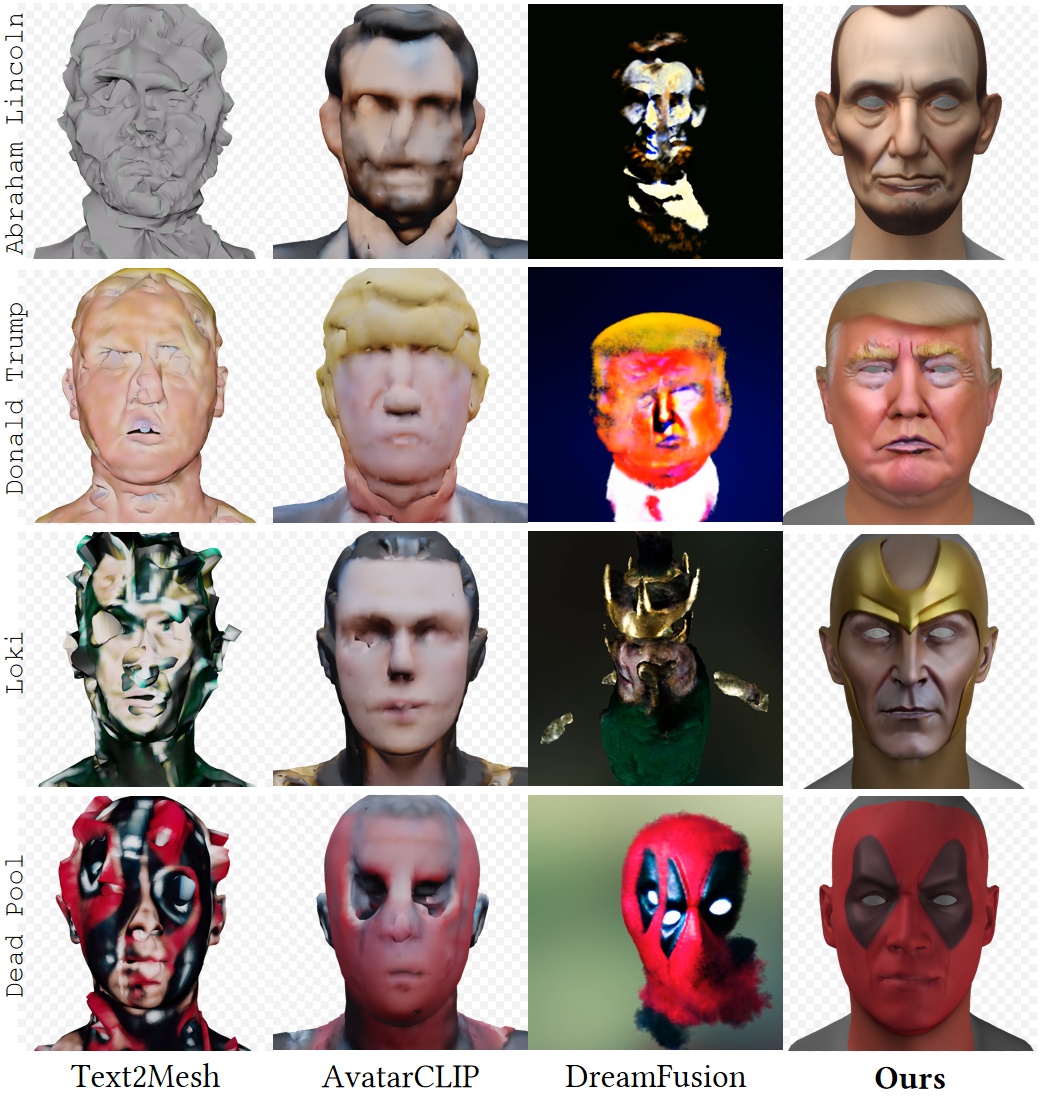}

\vspace{0.5cm}

    \begin{tabular}{ccc}\hline\hline
     \textbf{Method}    & \textbf{CLIP score} $\uparrow$ & \textbf{Running time}  $\downarrow$  \\\hline\hline
      Text2Mesh   &   0.2109  & $\sim 15$ mins\\\hline
      AvatarCLIP   &   0.2812  & $\sim 5$ hours\\\hline
      Stable-DreamFusion   &  0.2594 & $\sim 2.5$ hours \\\hline
      \textbf{Ours}   &   \textbf{0.2934} &  \textbf{$\mathbf{\sim 5}$ mins} \\\hline
    \end{tabular}

    \caption{
Qualitative and quantitative comparison between prompt controlled generation methods.
We show several generated results of each method (upper) and quantitative results (lower).
Compared to previous methods, our method preserves more details on appearance while keeping a fine-grained geometry. In addition, our method achieves the best text-matching score and has the least running time.
    }
    \label{fig:compare}
\end{figure}

\begin{figure}[t]
    \centering
    
    \begin{tabular}{cc}\hline\hline
     \textbf{Case}    &  \textbf{Resemble the specific character?}\\\hline\hline
      Celebrity   &   72.3\%   \\\hline
      Out of distribution   &   71.6\%  \\\hline
    \end{tabular}
    
\vspace{0.5cm}

    \includegraphics[width=\linewidth]{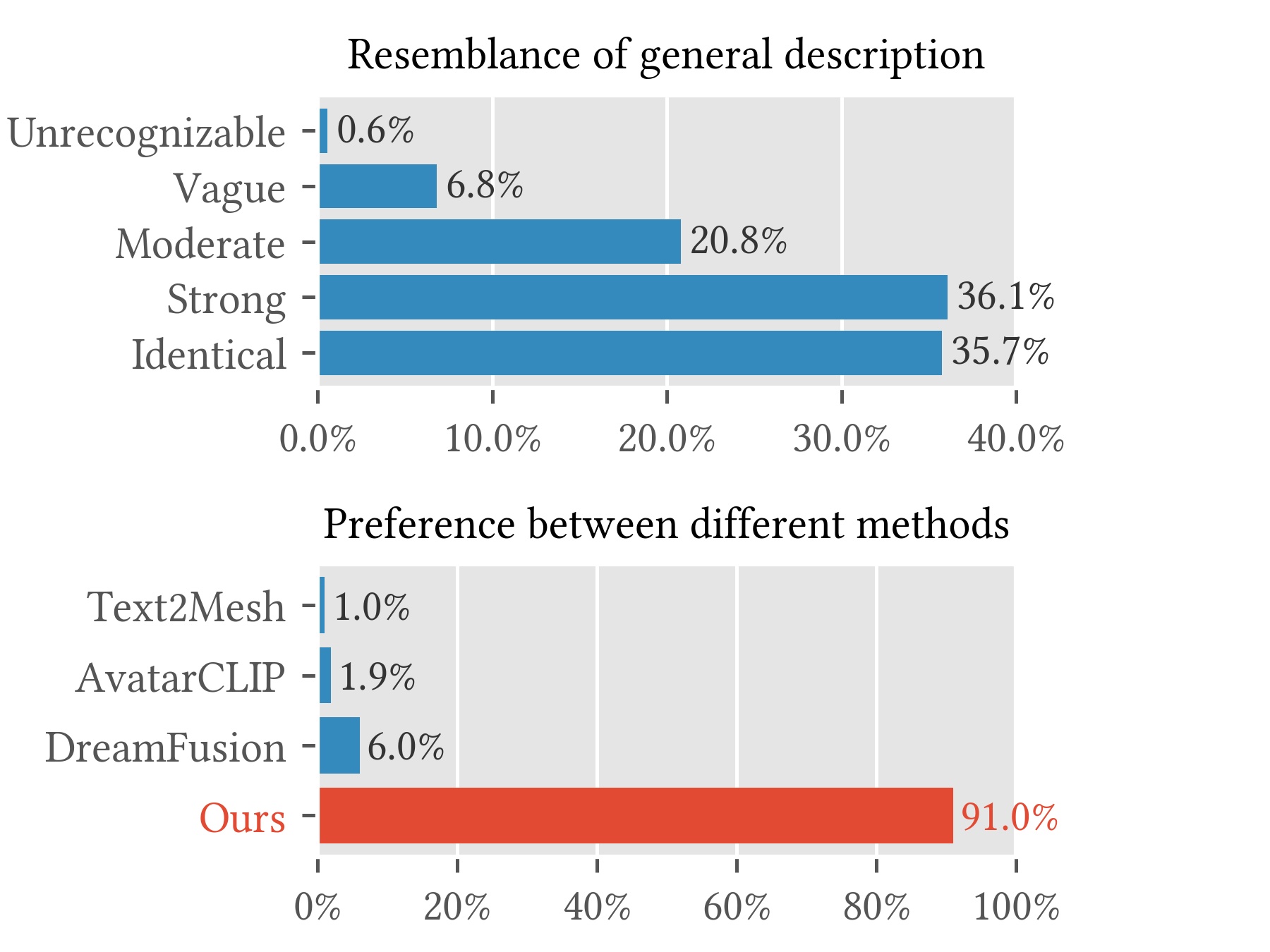}

    \caption{
Quantitative results of user study. The upper table illustrates our high resemblance to specific characters, while the lower figures demonstrate the high user ratings on matching general descriptions and the extremely high user preference for our method compared to other methods.
    }
    \label{fig:user}
\end{figure}


\section{Experiments}

In this section, we present the experiment results of DreamFace for generating animatable neural facial assets. 
We first introduce the implementation details and showcase a gallery of generated high-quality assets by DreamFace, highlighting the wide scope of applications that our approach enables. We then provide a detailed evaluation of modules in our pipeline, including geometry, appearance and animation modules, both qualitatively and quantitatively. 
At last, we compare with the state-of-the-art methods, followed by a comprehensive user study.

\subsection{Implementation details}

%
%
We rely on the generic LDM and texture LDM for geometry generation and texture diffusion. In our implementation, we use the pretrained LDM checkpoint~\textit{stable-diffusion-v1-5} from RunwayML~\cite{stable-diffusion-v1-5} as our generic LDM, and use the CLIP checkpoint \textit{clip-vit-large-patch14} from OpenAI~\cite{clip-vit-large-patch14}. While for the texture LDM, we duplicate a generic LDM and finetune it as described in~\ref{subsec:TextureLDM}. We train the texture LDM on two Nvidia A6000 GPUs for 150 epochs which takes about 12 hours.
%

During the geometry generation process, we sample one million candidates from ICT-FaceKit according to our pre-defined distribution and then perform 300 steps of detail carving. In the texture diffusion stage, we perform 200 steps of dual-path optimization in latent space and 200 steps in image space. By our two-stage design, the SDS-based generation process is very efficient, enabling the generation of a high-quality facial asset within 5 minutes on a single Nvidia A6000 GPU.

\subsection{Generation Results}

DreamFace is a novel framework for generating realistic 3D facial assets that are not only true-to-life in appearance, but also rich in animation capabilities. Our approach allows for the creation of highly detailed and personalized characters, from fashion icons to exotic creatures from fiction and film, all through the use of simple textual prompts.
%

By only providing a simple prompt of celebrity or general description, one can create realistic 3D facial assets that highly resemble the described characteristics, as demonstrated in Fig.~\ref{fig:gallery} and Fig.~\ref{fig:gallery2}, respectively. With DreamFace, one can even generate faces of fashion icons or unreal humanoid characters from fiction, movies, or even dreams in mind while retaining high recognition, as illustrated in Fig.~\ref{fig:outof}.
By leveraging both the generic LDM and the texture LDM, DreamFace generates animatable neural facial assets that are compatible with modern computer graphics pipelines and achieve photo-realistic rendering, utilizing the robust prior knowledge learned by these generative models through natural language prompts.
%

One can continue to customize its facial appearance for novel effects with additional textual guidance or even hand-make sketches, as illustrated in Fig.~\ref{fig:editing}. Our texture LDM directly supports denoising the existing texture under prompt guidance, such as adding wrinkles or makeup. By mixing the weight of trained texture LDM with a generic LDM, the mixed texture LDM is capable of generating desired patterns from noise or hand-make sketches in the masked areas that seamlessly blend with original diffuse maps (refer to stable-diffusion-webui~\cite{stable-diffusion-webui} for detail implementation).
%

Empowered by DreamFace, one can easily animate his creation using in-the-wild video footage directly with nuanced performance captured. 
Given a generated facial asset with geometry $\mathbf{T^\dagger}$, we directly inference its geometry from a input RGB image $\mathbf{V}$ by $\mathbf{\hat G}=\mathcal{G}_\text{geo}(\mathbf{T^\dagger},\mathcal{E}_\text{exp}(\mathbf{V}))$.
As illustrated in Fig.~\ref{fig:drive}, our animation pipeline empower animatability by generating personalized expressions and faithfully restoring the detailed performance.
With DreamFace, even novices can naturally create characters straight from their imagination, ready to be driven and rendered in stunning detail.

\subsection{Evaluation}
\subsubsection{Evaluation of texture LDM}
\label{sec:evaltextureldm}

In this section, we evaluate the quality of the generated textures produced by our texture LDM. Specifically, we employ Kernel Inception Distance (KID) to measure the performance of texture generation without classifier-free guidance on our full dataset and textures only from our desired domain.
To evaluate the performance of our approach, we conduct experiments with three different settings: \textbf{w/o m\&p}, \textbf{w/o prompt}, and \textbf{full}.
\textbf{w/o m\&p} denotes the naive LDM training pipeline without prompt tuning or masking. \textbf{w/o prompt} denotes the variant without prompt tuning, and \textbf{full} denotes our pipeline with both designs.
We generate 1000 samples for each setting and calculate KID 50 times with a batch size equal to the minimum available size of each pair. As illustrated in Fig.~\ref{fig:ablation}, \textbf{naive} results in lighting remaining and unwanted elements like hat and black area. \textbf{w/o prompt} removes the unwanted elements like black area but may produce textures with inconsistent skin tones. Our \textbf{full} generates desired textures with high quality, which demonstrates the effectiveness of our designs.

\subsubsection{Detail carving}

Detail carving, as proposed in Sec.~\ref{subsec:detailcarving}, is a key step of our framework that further reveals characteristics at the geometry level. Let \textbf{w/o carving} denote the coarse geometry and let \textbf{with carving} denote our detailed geometry. As illustrated in Fig.~\ref{fig:carving}, different carved geometries demonstrate the capability of our approach both in capturing characteristic traits of a specific individual and generating distinct facial features even across species.

\subsubsection{Ablation study of texture diffusion}
\label{subsucsec:abla}

Our proposed texture latent code generation pipeline, as outlined in Sec.\ref{subsec:latentspacesds}, utilizes both SDS in the latent space and joint optimization using a texture LDM to generate high-quality diffuse maps. 
Latent space SDS produces texture latent code more efficiently due to low resolution, while only using image space SDS takes much longer time (about 15 minutes, which is 3-4x times longer compared to latent space SDS) to generate satisfactory results.
To evaluate the performance of our pipeline, we compare it to two variations: one without the use of SDS in the latent space or texture LDM (similar to DreamFusion~\cite{poole2022dreamfusion}), denoted as \textbf{w/o l\&t}, and another one without the use of texture LDM (similar to Latent-NeRF ~\cite{metzer2022latent-nerf}), denoted as \textbf{w/o texture}. Our full pipeline, denoted as \textbf{full}, includes both SDS and texture LDM. As shown in Fig.\ref{fig:latent}, the results demonstrate that the \textbf{w/o l\&t} pipeline produces diffuse maps with a lot of noise and artifacts, \textbf{w/o texture} pipeline produces diffuse maps where components are drifting in UV space and lighting exists, while the \textbf{full} pipeline gives the best results that meet the standard used in physically-based rendering.

\subsubsection{Texture augmentation}
\label{subsec:textureaug}
We demonstrate the effectiveness of texture augmentation in Fig.~\ref{fig:superres}.
After generating satisfactory texture latent code, our texture translation and augmentation module further provide detailed physically-based textures, including the diffuse map, specularity map and normal map in high resolution, which are essential for photo-realistic rendering.

\subsubsection{Comparison of animation generation}

We compare our animation method that generates personalized expressions from images to the previous method DECA \cite{feng2021deca} that infers parameters of general expression blendshapes. To adapt FLAME \cite{li2017flame} blendshapes to our asset, we transfer the expression parameters (blendshape weights and jaw rotation) to control our generated facial asset with the rig that same as FLAME. 
Let \textbf{DECA}, \textbf{Ours-D}, \textbf{Ours} denote the original results from DECA, the transferred expression using our asset, and our enhanced animation scheme, respectively. 
As illustrated in Fig.~\ref{fig:deca}, compared to DECA, our framework provides personalized expressions and more accurately restores the detailed expression in the video. 

\subsubsection{Comparison with state-of-the-art methods}

We further present a comprehensive comparison of our proposed method with state-of-the-art prompt-controlled generation techniques, including Text2Mesh \cite{michel2022text2mesh}, AvatarCLIP \cite{hone2022avatarclip} and DreamFusion \cite{poole2022dreamfusion}. Our goal is to evaluate the quality and efficiency of our approach in comparison to other methods.
To conduct the comparison, we generate 10 different characters using each method. For Text2Mesh, we use the official implementation. For AvatarCLIP, we use the generated results from the official website. For DreamFusion, we use a re-implementation version \cite{stable-dreamfusion} that uses Stable Diffusion. For all these approaches, we use the prompt "the realistic face of SOMEONE" for the generation.
To quantitatively evaluate the generated results, we calculate the CLIP score by computing the cosine similarity of image features and text features using the prompt ``the realistic face of SOMEONE''. Additionally, we also measure the running time of each method.
As illustrated in Fig.~\ref{fig:compare}, our method demonstrates its capability to generate high-quality realistic assets efficiently and outperforms others in terms of both CLIP score and running time.

\subsubsection{User study}

Finally, we conducted a comprehensive user study to evaluate the performance of our generated facial assets in terms of their ability to match the given prompts and their level of resemblance to specific characters or general descriptions. We recruited 180 volunteers to participate in the study, which consisted of three main evaluations: matching specific characters, matching general descriptions, and user preference across different methods.
For the evaluation of specific characters, we generated 27 different samples that included both celebrities and out-of-distribution characters and asked volunteers to rate the level of the resemblance of the generated results to the specific characters. For the evaluation of general descriptions, we generated 10 different cases with diversity and asked volunteers to score the conformity of the results with the descriptions on a five-point scale. Additionally, we compared our method with Text2Mesh, AvatarCLIP, and DreamFusion using 10 different prompts, and asked volunteers to select their preferred method.
As shown in Fig.~\ref{fig:user}, our method achieved high levels of resemblance and is significantly more preferred than the other three methods.

\subsection{Limitation}


As a brave attempt, DreamFace presents a novel approach for the progressive generation, editing, and animation of neural facial assets using simple text prompts. Our framework demonstrates a high level of realism and resemblance in the generated assets, even for novice users. However, it is important to note that there are limitations to our approach.
One limitation is that while we have incorporated hairstyle generation into our framework, the generation of full facial components such as eyes and mouth interiors is currently not possible. The modeling and rendering of eyes, in particular, presents significant technical challenges that are yet to be fully addressed.
Another limitation is that as a framework based on prompt-conditioned diffusion models, DreamFace is constrained by the capabilities of these models. For example, while GAN-based methods have the ability to invert a given sample easily, the inversion of diffusion models has not been fully explored. While the performance of DreamFace is influenced by the state-of-the-art in diffusion model research, further advancements in this area may improve our results.
Lastly, while our framework has demonstrated strong animation capabilities through the native support of blendshapes and an enhanced animation scheme, there is still potential for further research in the area of prompt-based animation control. The generation of facial motion, including lively expressions and nuanced performance, is a particularly interesting and challenging problem that remains to be solved.

\paragraph{Potential ethical implications}

As a text-driven generation method, DreamFace operates on the output of pre-trained large-scale vision-language models such as CLIP and Stable Diffusion, and is subject to any biases present in their training data, like gender or racial preferences. However, it is essential to note that these biases are not unique to DreamFace, and are inherent in any generation method that relies on these pre-trained models.
Additionally, the ease of use and high quality of the generated assets may also raise concerns about the potential for misuse, such as creating fake videos or impersonating individuals.
It is important for future work to consider and address these ethical issues in the development of text-driven generation methods by developing methods to mitigate biases in pre-trained large-scale vision-language models and ensuring that the data used for training these models are diverse, representative, and carefully reviewed.

\section{Conclusion}
We have presented DreamFace, to progressively generate personalized 3D faces that are compatible with CG engines from only text-prompt controls. Through DreamFace, even novices can naturally create 3D human characters with desired shapes and textures they have in mind, easily customize the creations for novel effects like aging and virtual makeup, and even further animate the creations using in-the-wild video footage. 
Specifically, our coarse-to-fine scheme efficiently generates high-quality geometry, including the coarse geometry with a unified topology and the nuanced displacement and normal details. For appearance generation, our dual-path mechanism organically combines two latent diffusion models, i.e., a generic LDM and a texture LDM, achieving diverse and consistent results with the textural specification. Our two-stage optimization in both the latent and image spaces achieves highly efficient and fine-grained synthesis, enabling the mapping from the compact latent space to physically-based textures. Our neutral assets naturally support blendshapes-based facial animations, while the accompanied neural animation scheme further provides personalized fine-grained animation from only video input.  
Extensive experimental results and user studies have demonstrated the effectiveness of DreamFace. We believe that our approach renews the generation of accessible 3D facial assets with physically-based rendering quality and rich animation ability in the neural and prompt-interaction era. It not only benefits the CG production industry but also incentivizes numerous innovative applications for VR/AR and the emerging Metaverse.




\bibliographystyle{ACM-Reference-Format}
\bibliography{sample-bibliography}

\end{document}